# How hard can it be? Quantifying MITRE attack campaigns with attack trees and cATM logic

STEFANO M. NICOLETTI, University of Twente, the Netherlands
MILAN LOPUHAÄ-ZWAKENBERG, University of Twente, the Netherlands
MARIËLLE STOELINGA, University of Twente and Radboud Universiteit, the Netherlands
FABIO MASSACCI, Vrije Universiteit, the Netherlands and University of Trento, Italy
CARLOS E. BUDDE, Technical University of Dennmark, Denmark

In the cyber threats landscape, Advanced Persistent Threats carry out attack campaigns—e.g. operations Dream Job, Wocao, and WannaCry—against which cybersecurity practitioners must defend. To prioritise which of these to defend against, experts must be equipped with the ability to evaluate the most threatening ones: they would strongly benefit from (a) an estimation of the likelihood values for each attack recorded in the wild, and (b) transparently operationalising these values to compare campaigns quantitatively. Here we construct such a framework: (1) quantifying the likelihood of attack campaigns via data-driven procedures on the MITRE knowledge-base, (2) introducing a methodology for automatic modelling of MITRE intelligence data, that captures any attack campaign via template attack tree models, and (3) proposing an open-source tool to perform these comparisons based on the cATM logic. Finally, we quantify the likelihood of all MITRE Enterprise campaigns, and compare the likelihood of the Wocao and Dream Job MITRE campaigns—generated with our proposed approach—against manually-built attack tree models. We demonstrate how our methodology is substantially lighter in modelling effort, and capable of capturing all the quantitative relevant data. To ensure broader applicability, further validation with cybersecurity experts is recommended, especially by sourcing more manually-built models.

## 1 INTRODUCTION

The modern threat landscape sees cybersecurity practitioners facing new menaces every day. Organized in threat groups—such as the Lazarus, HAFNIUM and Cobalt Groups—threat actors configure themselves as Advanced Persistent Threats (APTs) that carry out systematic attack campaigns. Stark examples of such organized attacks are operations Dream Job, Wocao, WannaCry or the SolarWinds Compromise, to name a few. In the face of the overwhelming number of organized

Authors' addresses: Stefano M. Nicoletti, s.m.nicoletti@utwente.nl, University of Twente, Enschede, the Netherlands; Milan Lopuhaä-Zwakenberg, University of Twente, Enschede, the Netherlands; Mariëlle Stoelinga, University of Twente, Enschede, and Radboud Universiteit, Nijmegen, the Netherlands; Fabio Massacci, Vrije Universiteit, Amsterdam, the Netherlands and University of Trento, Trento, Italy; Carlos E. Budde, cesbu@dtu.dk, Technical University of Dennmark, Lyngby, Denmark.

campaigns, it is paramount that security practitioners possess effective instruments for transparent and accountable decision making. This is particularly important in order to evaluate which risks are most threatening, and which campaigns to prioritize against when defending.

*A starting point: the MITRE ATT&CK knowledge base.* MITRE ATT&CK is a useful resource for campaign analysis and quantification, specially for custom applied studies of cybersecurity practitioners. Their public-access knowledge base receives weekly updates in the STIX Data github. Moreover, the main website at attack.mitre.org offers a high-level perspective with in-depth documentation and links to tools like the MITRE ATT&CK Navigator and Python utilities at http://attack.mitre.org/resources/attack-data-and-tools/. MITRE ATT&CK is today the *de facto* standard knowledge base to register and refer to campaigns that ocurred in the wild.

*Problem statement.* In spite of its status and usefulness, MITRE ATT&CK does not present a straightforward way for the quantitative comparison of attack campaigns, especially when considering their likelihood. More specifically:

(1) there is currently no granular and flexible way to carve likelihood data from MITRE, in order to easily evaluate the probability of campaigns happening and
(2) there is no catch-all method to formally model and query any possible campaign (to be) recorded in MITRE, to favour structured analysis and comparison of quantitative properties of those attack campaigns.

Having a method that caters for these gaps would help cybersecurity practitioners when distributing resources to defend against APTs. This benefits both low-level technical knowledge with quantitative data based on MITRE recordings, as well as high-level decision making on where to spend how many resources.

*Our contribution.* To address the above, we propose a framework grounded in data mined from MITRE public resources. The framework can quantify likelihood data from attack techniques, to be used in formal models for quantitative comparisons among campaigns that use the techniques.

More specifically, we propose a systematic way to carve probabilistic data from MITRE ATT&CK. This is instrumental to estimate the likelihood of observing a specific attack technique to achieve an attacker's tactical goal[†]. We further operationalize this via model templates that are constructed automatically from MITRE ATT&CK data: these models represent MITRE campaigns, and allow computations of the probability to succeed in executing such campaign. For this last step, cATM underpins an automatic framework for campaign generation and comparison via a formal logic, that revises the ATM logic [34] for cybersecurity purposes.

In particular, the model templates can be generated in three difficulty levels, encoding the hardness of an attack from the perspective of a threat actor: hard, easy, and default. They are constructed using the popular *attack trees* (ATs) [41] formalism and provide a baseline, against which to compare custom AT models that can be generated *ad hoc* by field experts. Furthermore, these templates allow for cross-campaign comparison, empowering practitioners with tools to assess relative likelihoods of attack campaigns.

Finally, we package this pipeline into a publicly-available artifact, and automatically generate AT templates for every campaign recorded in MITRE ATT&CK. Moreover, we validate our baseline motivation by constructing *ad hoc* models for the Dream Job and Wocao campaigns, and compare them to our automatically generated AT templates.

We offer a visual representation of our contributions, as well as a detailed summary, in Fig. 1.

[†] MITRE does not distinguish the success rate of attack techniques in deployed systems, so neither can our framework. Technical reports like CISA RVA [7] do this partially: Sec. 8 discusses the incorporation of such statistics in our framework.

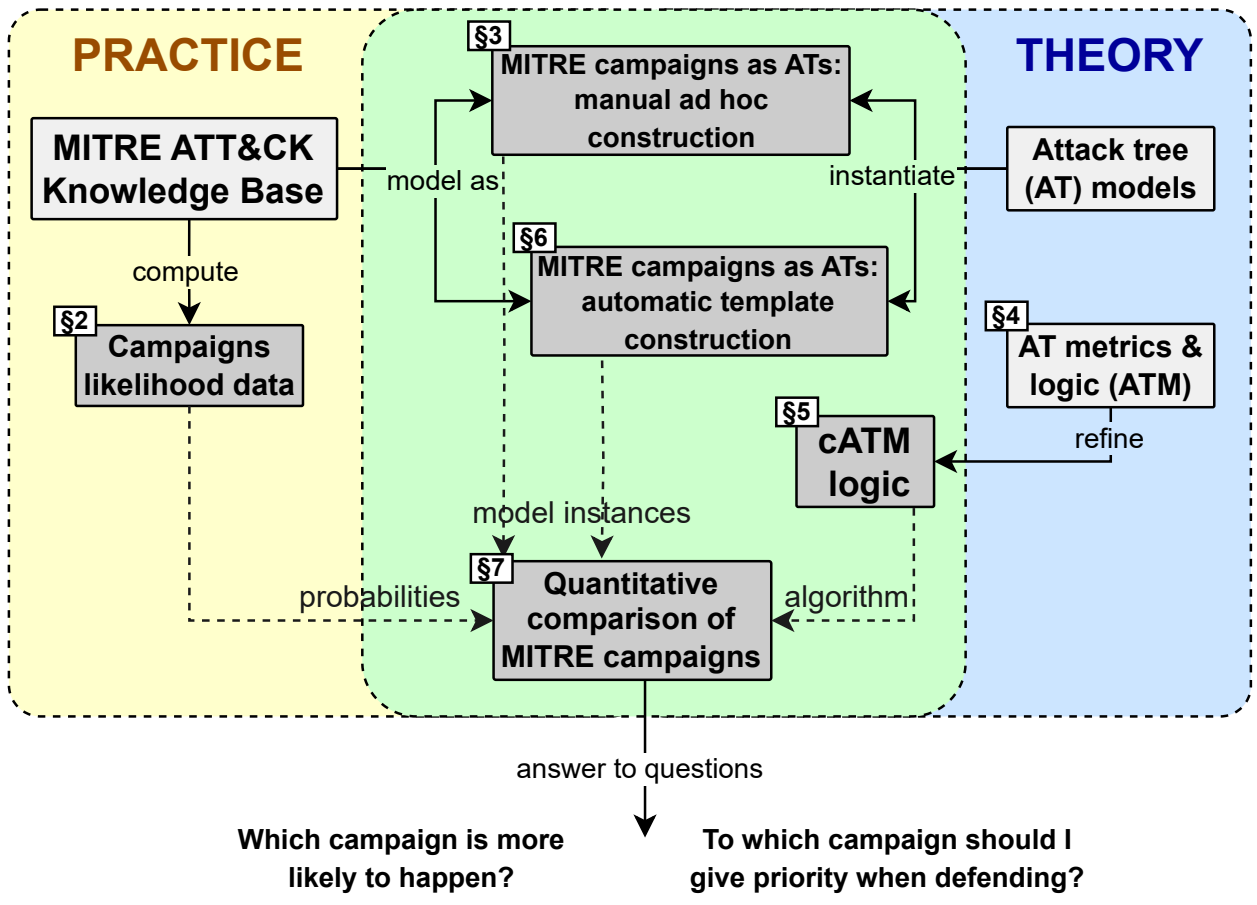

Fig. 1. A diagram detailing the structure the manuscript, where our contributions are the darker rectangles.

**Summary of contributions:**

1. We quantify likelihood of attack campaigns via data-driven procedures **on the MITRE knowledge base**:
   (a) **§ 2.2**: **computing probability-like values (*likelihoods*)** for the frequency with which a technique was used to achieve each tactical goal.
   (b) **§ 4.4**: introducing a **security index** (an order-inverting transformation of the likelihood) as a data-driven quantity to compare the likelihood of observing MITRE campaigns.
2. **§ 3**: We showcase how **individual MITRE campaigns can be manually modelled as an attack tree**, and present two instances that exemplify this procedure for the Dream Job and Wocao campaigns.
3. We revise **the AT logic from** [34] for cybersecurity practice and for comparative use on MITRE campaigns, concretely by:
   (a) **§ 4.3**: **adding support for interval values**, allowing for modelling uncertainty.
   (b) **§ 5**: **minimising it from 4 to 2 production layers**, keeping only the constructs that are essential to compute quantitative metric values of an attack.
4. **§ 6**: We introduce **automatically-built attack tree templates that model MITRE intelligence data**. These are built from public data, via a methodology that is *complete* in the sense that it captures any attack campaign including all registered attack techniques and tactics recorded in the MITRE ATT&CK knowledge base.
5. **§ 7**: We validate our approach by **comparing the likelihood of the Wocao (C0014) and Dream Job (C0022) MITRE campaigns** – generated with our proposed methodology – against "ad hoc" traditionally-built ATs, demonstrating how our methodology is substantially lighter in modelling effort, and still capable of capturing all the quantitative relevant data.
6. **§ 7.1**: We provide a publicly accessible repository that stores an artifact, packaging our contribution in an easy-to-use implementation, at DOI 10.5281/zenodo.14193936.

*Structure of the paper.* We showcase how to quantify likelihood of campaigns via data-driven methods on MITRE in Sec. 2, we further propose a formal encoding of MITRE campaigns via attack tree models in Sec. 3 and showcase traditionally-built ATs for the Dream Job and Wocao campaigns, we offer a light-effort framework for quantification and comparison of MITRE campaigns exploiting a formal logic (Sec. 5) and automatically generated AT templates (Sec. 6) and we validate our approach and perform comparisons on the Wocao and DreamJob campaigns – both on *ad hoc* and template ATs– in Sec. 7. We reflect on future perspectives and conclude the paper in Sec. 8.

## 1.1 Related work

A body of work exists on the use of MITRE for the analysis of cyber attacks. While surveys like [17] point at many research approaches that build ATs automatically from data, these usually mine process information or event logs from which they infer an AT. Two relevant examples are [14, 38], which use MITRE qualitative data such as Common Vulnerability and Exposures (CVE) and Common Weakness Enumeration (CWE) identifiers in their nodes, or even numeric information such as the Common Vulnerability Scoring System (CVSS) or the Exploit Prediction Scoring System (EPSS) of a vulnerability identified by a CVE.

While such approaches may include quantitative information in the models that they generate, their main purpose is to aid (mostly by semi-automatic procedures) the *generation* of an AT, as opposed to its use in the *computation* of a quantity from the generated model. In contrast, for us, since MITRE has hierarchically-structured information that is amenable to quantification, we leverage it to create ATs in a fully-automatic manner via a deterministic process. We then exploit these models via the cATM logic, to compute values inherent to each campaign, hence allowing a quantitative comparison among them.

All in all, while we are more closely related to studies that propose quantifications of either the probability or impact of an attack, we also share motivations, or certain aspects of our approach, with many other recent studies. A classification of such works in terms of research proximity is:

1. studies that use MITRE ATT&CK to quantify cyber attacks: [15, 16, 48]—closest to us;
2. studies that use MITRE ATT&CK for qualitative analysis of cyber attacks: [1, 2, 22];
3. studies that quantify cyber attacks without using MITRE ATT&CK: [34, 37, 47];
4. exploitation of MITRE ATT&CK for extrinsic empirical analyses: [21, 23, 43, 45];
5. other related works that do not fit the above categories: [5, 32].

#### 1.1.1 *Works that quantify attacks using MITRE ATT&CK.*

Xiong et al. [48] propose a threat modeling language, *enterpriseLang*, based on MAL [15] and the MITRE ATT&CK Matrix. This is used to model IT systems and enable attack simulations, and it is demonstrated on the emulation of MITRE campaigns. While the application of [48] is not focused on quantitative analysis, their end-result could in principle be labelled with values that represent externally acquired quantitative data of the modelled attacks. Instead, Kim et al. [16] introduce custom quantifications for MITRE ATT&CK campaigns, using the MITRE Matrix and Lockheed Martin Cyber Kill-Chain. For MITRE, each tactic—12 out of the 14 tactics recognised today—of each campaign is assigned a score equal to the number of techniques used in it. This scoring approach is available today in the MITRE ATT&CK Navigator, and disregards nuances such as data granularity, e.g. the simultaneous marking of a technique and some of its subtechniques in a MITRE ATT&CK campaign.

In contrast to these works, our approach is adaptable to different data granularity levels (see Sec. 2.2.3), and it is designed for the quantitative analysis of any campaign—already recorded or to come—by taking as input data directly available in MITRE.

*1.1.2 Works that are centred on non-quantitative aspects of MITRE ATT&CK.* Since its release in 2018, MITRE ATT&CK has become an authoritative source of cybersecurity intelligence used in many scientific works [1, 2, 10, 22, 39, 40]. From that literature we highlight the following, which are closest in nature to our approach. Lee and Choi [22] vectorise MITRE campaigns data to recognise the action of the corresponding attack groups, and Akbar et al. [1] resort to data mining and an LLM to propose relevant defensive actions based on user-identified attacks. In turn, Akbar et al. [2] introduce an ontology that describes MITRE ATT&CK components, to facilitate the recognition of ongoing campaigns by security analysts.

These works share the same study focus as our research, i.e. the enhancement of MITRE data for systematic analyses of cybersecurity attacks—however, they develop strategies that could best be described as qualitative in nature, in contrast with our quantitative focus.

*1.1.3 Works that quantify cybersecurity data without resorting to MITRE ATT&CK.* Woods and Böhme [47] introduce a causal model inspired by structural equation modeling, that explains cyber-risk outcomes in terms of latent factors. This model is then used to classify empirical cyber-harm and -mitigation studies—see tables 1 and 3 therein. Instead, from suspicious system events, Patil et al. [37] generate a provenance graph based on advanced persistent threats (APTs) campaigns that match them, to classify—probabilistically—the ongoing attack as a known APT campaign or not. Finally, Nicoletti et al. [34] propose a first-order logic to query quantitative data from attack campaigns (represented as attack trees), including probability and time of attack.

In contrast to our research, the above propose general theoretical frameworks that do not consider MITRE. However, we do align with many of the theoretical developments of [34], which we adapt and refine to fit data as provided in MITRE ATT&CK.

*1.1.4 Works that exploit (but do not build upon) MITRE ATT&CK.* Lee et al. [23] propose correlation analyses on data traffic, that use cross-Tactic correlation to detect exfiltration as described by MITRE. This is used to differentiate benign traffic from exfiltration activity by APTs. Shin et al. [43] organise MITRE ATT&CK data into a framework, that they use to identify countermeasures against APTs' phishing campaigns. Tsakoulis et al. [45] characterise malicious behaviour in Linux OS execution flow by association to MITRE data. They employ this methodology to emulate SysJoker malware via the exploit attack graph from MITRE ATT&CK. Finally, Kumar and Thing [21] generate test-cases for MITRE lateral-movement detection of a new technology proposed for 5G networks.

All these works leverage information available in MITRE ATT&CK, and use it for extrinsic empirical ends such as traffic analysis and countermeasure identification. In contrast, our approach reflects back on MITRE by enriching its data with quantitative information—see Sec. 2.2.

*1.1.5 Other related studies.* Munaiah et al. [32] codify events, extracted from penetration testing competitions, as MITRE techniques and tactics. Their results demonstrate the usability of MITRE ATT&CK as a grounding for cybersecurity activities. In contrast, Skjøtskift et al. [44] use MITRE ATT&CK to structure semantic models, namely relations between MITRE techniques and the actions that a threat-actor needs to perform in order to apply the technique, to answer questions such as "*What did most likely happen prior to this observation? and What are the adversary's most likely next steps given this observation?*" [44]. Similarly, Bleiman et al. [5] showcase how social engineering practices can be mapped onto the MITRE ATT&CK framework, via an academic course study that included a MITRE representative.

## 2 MITRE INTELLIGENCE DATA AND LIKELIHOOD VALUES

We structure MITRE intelligence data, gathered by and available to security practitioners, in a manner that enables quantitative comparisons of likelihood among attacks.

### 2.1 The MITRE ATT&CK knowledge base

Our models and our approach are based on MITRE ATT&CK, which is a "*knowledge base of adversary tactics and techniques based on real-world observations [...] used as a foundation for the development of specific threat models and methodologies in the private sector, in government, and in the cybersecurity product and service community.*" [28]

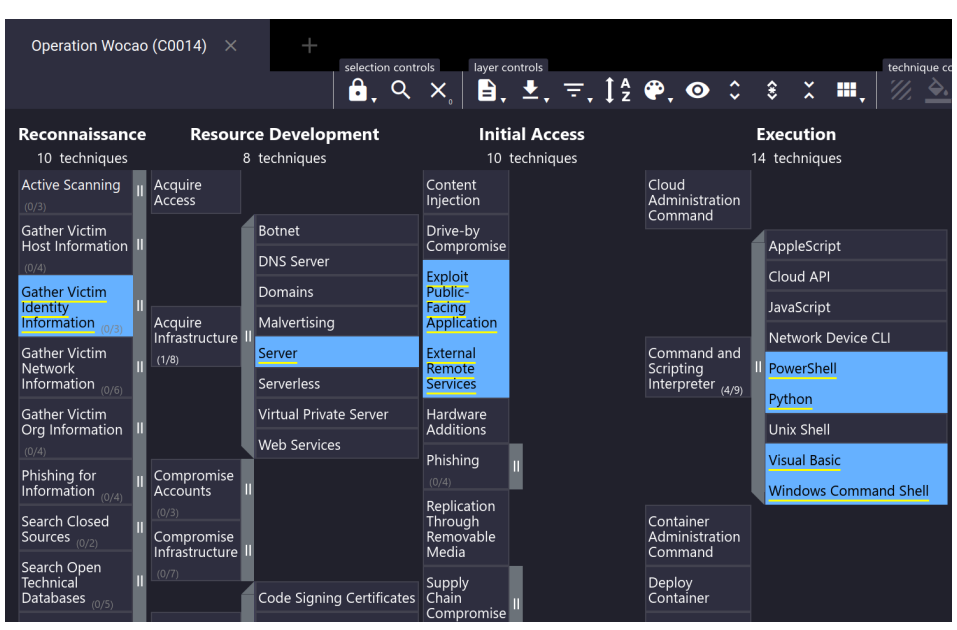

Fig. 2. MITRE ATT&CK campaign navigator
https://mitre-attack.github.io/attack-navigator

#### 2.1.1 *Tactics.*

MITRE defines a standard cyberattack structure, consisting on a partially-ordered set of *tactics* that represent the adversary's tactical goal—see [30]. Examples include reconnaissance (MITRE ID TA0043) and exfiltration (TA0010). The former represents the intentions of an adversary who is trying to gather information for use in planning future operations. Naturally, this is one of the first tactics that an adversary aims for when initiating an attack campaign—in contrast, exfiltration is one of the last. In fact, exfiltration represents the goal to steal (as in remove) data from the target network, for instance by compressing, encrypting, and sending the data to an external malicious server.

#### 2.1.2 *Techniques.*

In these examples, compression and encryption are concrete actions that the adversary executes to achieve tactical goals such as exfiltration. These actions are encoded in MITRE as *techniques*, and they describe the technical details of an adversary's actions to achieve a tactic—see [31]. For example, MITRE technique T1052.001 (*Exfiltration Over Physical Medium: Exfiltration over USB*) represents an attacker's attempting "*to exfiltrate data over a USB connected physical device [...] The USB device could be used as the final exfiltration point or to hop between otherwise disconnected systems*" [28].

#### 2.1.3 *Campaigns.*

The collection of tactics and techniques available in MITRE ATT&CK were gathered from real-world observations by security experts. This intelligence on intrusion activity is recorded in MITRE as cyberattacks known as *campaigns*. For example, Wocao (C0014) was a China-based cyber-espionage campaign that targeted several organizations around the world, compromising services such as aviation, construction, energy, finance, etc. [29]. MITRE ATT&CK provides an *Attack Navigator* (https://mitre-attack.github.io/attack-navigator/): a graphical overview tool to represent techniques employed to achieve tactical goals in any campaigns—see Fig. 2 for an excerpt of the navigator view for the Wocao campaign.

*Formalisation.* Data in the MITRE knowledge base is kept and updated in a structured but informal manner, with natural language descriptions (e.g. technical reports) and an interface that is updated as new information is acquired. In other words, MITRE ATT&CK intelligence data is not formal, and therefore retrieving it in a systematic way—e.g. to model it formally as attack trees—is difficult:

> MITRE ATT&CK data is not formal, hindering its rigorous, unambiguous, and quantified modelling. We show how such formalisation, including probability quantifications, can be done.

*Likelihood and probability values.* Throughout this article we write *likelihood* to indicate the chance of observing a real-world event, such as the prospect of suffering an attack in the next $x$ days, or that a concrete technique $E$ is used in such attack. These values are unknowable a priori, and can only be approximated based on past evidence and expert technical knowledge. In contrast, by *probability* we mean a concrete value $0 \leq p \leq 1$ that can be computed from existent data.

The absence of correct and complete information from the real world speaks against using raw probability values as uncontextualised approximations of likelihoods. Notwithstanding, when treated formally within a rigid framework, probabilities can still be extremely useful to compare odds and aid decision-making at design time. We expand on this next.

## 2.2 Probability metrics from MITRE data

*2.2.1 Campaigns, tactics, techniques.* MITRE campaigns provide statistics of the frequency with which threat actors use an attack technique. Using cybersecurity intel to forecast expected future activity—as is common practice [24, 49]—allows to interpret the frequency of a technique as its probability of occurrence.

Naturally, interpreting MITRE techniques' frequency as probabilities is not a direct measure of the likelihood of observing an attack in the real world. Besides the particular setting of the victims involved, this is due to the absence of perfect information, since the only data available comes from observable cyber-criminal activities that have been made public.

Nevertheless, the 600 techniques recorded in MITRE give some indication of likelihood, ordinal if not cardinal. To see this, consider a technique $E$ (e.g. where `/etc/passwd` was vulnerated) that was used in 77% of the campaigns to achieve tactic $A$ "gain credential access". Say further that technique $E' \neq E$ was used in 2% of the campaigns to achieve $A$. Then the expectation is that attacks for $A$ via $E$ are more likely than those involving $E'$, which companies can use to prioritise security measures—see e.g. the Threat Intelligence team at Würth Phoenix https://www.wuerth-phoenix.com/.

*2.2.2 Probabilities estimation.* We use the above to estimate, for each technique $E \in \mathcal{E}$ and tactic $A \in \mathcal{A}$ in MITRE, the probability $p_{E|A} = Prob(E \mid A)$ defined as the frequency with which $E$ was used with respect to all other techniques to achieve tactic $A$. To compute these values let $C[A] \in 2^{\mathcal{E}}$ be the set of techniques $\{E_1, E_2, \ldots\}$ used in campaign $C \in \mathcal{C}$ for tactic $A$, and let $A\upharpoonright$ be the multiset of techniques used for tactic $A$ across all campaigns, i.e. $A\upharpoonright = \biguplus_{C \in \mathcal{C}} C[A]$, where $\uplus$ denotes multiset union. Then the probability sought is given by:

$$p_{E|A} = Prob(E \mid A) \doteq \frac{\sum_{C \in \mathcal{C}} \mathbf{1}_{C[A]}(E)}{|A\upharpoonright|} = \frac{\text{how many } E'{=}E}{\text{how many } E'} \tag{1}$$

*Example 2.1.* Let $C_1, C_2$ be MITRE campaigns wherein tactics $A_1, A_2$ were achieved by threat actors as follows: $C_1 = \langle A_1{:}\{E_1\}, A_2{:}\{E_2\}\rangle$, and $C_2 = \langle A_1{:}\{E_1, E_2\}, A_2{:}\{E_3\}\rangle$. Then $C_1[A_1] = \{E_1\}$ and $C_2[A_1] = \{E_1, E_2\}$, and moreover $A_1\upharpoonright = \{\!\{E_1, E_1, E_2\}\!\}$ has cardinality $|A_1\upharpoonright| = 3$. Finally we compute the probability estimates: $p_{E_1|A_1} = {}^2/_3$, and $p_{E_2|A_1} = {}^1/_3$, and $p_{E_3|A_1} = 0$.

*Dempster-Shafer theory.* Note that we combine the data of different campaigns in a frequentist manner, as opposed to the more intricate ways of combining evidence of Dempster-Shafer theory [25, 42]. That theory expresses not only the aggregate opinion of multiple sources, but also the uncertainty due to disagreeing sources. In our case, every source is a campaign $C$, and its input on $p_{E|A}$ is binary: either $E \in C[A]$ or not. Dempster-Shafer theory does not allow the combination of evidence that is in such direct disagreement (which is the only one that MITRE currently provides).

*2.2.3 Data counting and use.* For most techniques, MITRE lists possible subtechniques through which the technique can be performed. For instance, $E$ = `System Services` (MITRE codename `T1569`) can be done via `Launchctl` (`T1569.001`) or `Service Execution` (`T1569.002`). Thus, we use Eq. (1) for subtechniques—rather than techniques—to obtain finer-grained information about the attack.

However, the data available across campaigns in MITRE is of different granularity, and in some cases only the high-level technique is known. Therefore, to normalise this data for subtechnique frequency counting, we consider the following cases that can be found for an arbitrary tactic $A$ in campaign $C$, concerning some high-level technique $E$ and its subtechniques:

- **Fine-grained data** (optimal case): if $C[A]$ contains only subtechniques $E_1, \ldots, E_n$ of a given high-level technique $E$, then count each $E_i$ as +1 occurrence of itself;
- **Coarse-grained data** (least knowledge): if $C[A]$ contains no data about subtechniques of $E$, but it does mention $E$, then count +1 to each subtechnique $E_i$ that theoretically exists for $E$—this is an unbiased worse-case scenario corresponding to an attacker trying all possible actions;
- **Mixed data**: if $C[A]$ contains subtechniques $E_1, \ldots, E_n$ of the high-level technique $E$, *and it also lists E as a specific technique used for A*, then use the fine-grained information by counting +1 for each subtechnique $E_1, \ldots, E_n$.

*Data sources.* To compute probabilities using eq. (1), the data available per campaign must distinguish the tactics for which each (sub)technique was employed. MITRE offers several resources, such as STIX https://github.com/mitre-attack/attack-stix-data/, and `mitreattack-python` https://github.com/mitre-attack/mitreattack-python/. From the various options, we found the required division of techniques per tactics (only) in the MITRE ATT&CK navigator https://mitre-attack.github.io/attack-navigator. In particular, this resource provides an overlay called `aggregate scores`, whose `sum` configuration counts the number of (sub)techniques used for each tactic. While that goes in the same direction than the data counting proposed above, it does not distinguish the different granularity cases—for instance, in the mixed-data case of `T1078` and `T1078.002` mentioned next, it provides a score of $1 + 1 = 2$. Thus, we use scripts to process the public data provided by MITRE ATT&CK using eq. (1) on the data counted as indicated in this section.

*Mixed data in MITRE.* The redundancy stemming from *mixed data* cases may be caused by the data-processing scripts that generate the public knowledge base from technical reports. We observe this e.g. for the Wocao campaign in MITRE ATT&CK [29], where *Valid Accounts* (`T1078`) coexists with *Domain Accounts* (`T1078.002`) for tactic *Initial Access*. This may be caused by the wording used in §§ 6.3.2 and 6.3.3 of the technical report, which mention the use of "*compromised VPN credentials*" and in particular "*Windows domain credentials*" [46]—as it happens, the former is part of the MITRE description of `T1078`, and the latter of `T1078.002`. Thus, in such cases we omit the redundancy and use the finer-grained information available, namely the subtechnique(s).

## 3 MITRE CAMPAIGNS AS (MANUAL) ATTACK TREE INSTANCES

An attack tree (AT) is a hierarchical diagram describing potential attacks on a system. Its root represents the attacker's goal, and the leaves represent basic attack steps (unrefined attacker actions). Intermediate nodes are labeled with gates, that determine how their children activate them. The most basic ATs have OR and AND gates; extensions exist to model more elaborate attacks.

*MITRE campaigns as ATs.* An individual MITRE campaign can be mapped to an AT instance, using leaves to represent techniques, and gates to represent the tactics for which they were used. While ad hoc decisions may be needed to instantiate this procedure for each campaign, a natural map exists, whereby the nodes of the AT instance come from the empirical evidence that is available to cybersecurity practitioners in MITRE. These models support attributes (such as probability values), that can be used for quantitative analyses—we exploit this in the following sections.

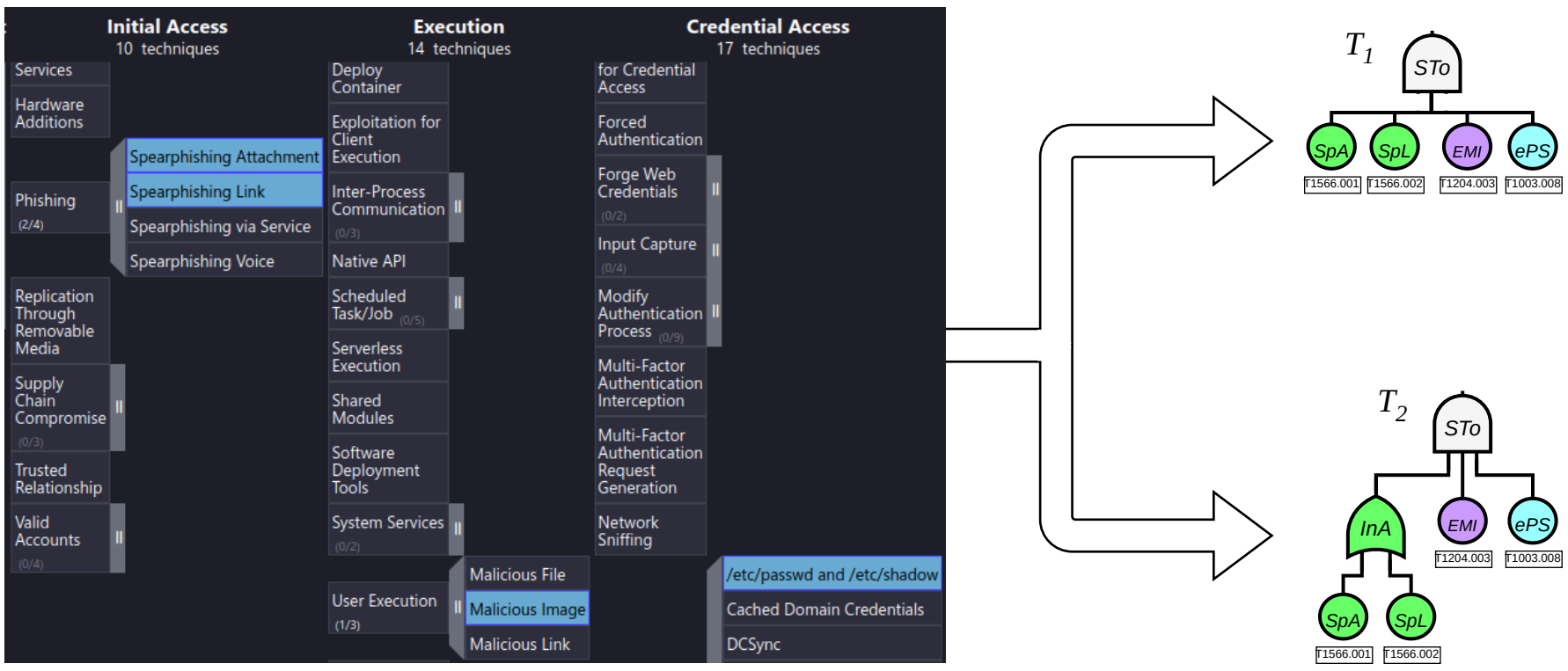

Fig. 3. Toy MITRE campaign represented as (different) attack trees

For a concrete example consider Fig. 3 (page 9) which depicts a very basic campaign composed only of three tactical goals: *Initial Access*, *Execution*, and *Credential Access*. These are achieved by leveraging four techniques, namely *Spearphishing-Attachment* and *-Link*, *Malicious Image*, and */etc/passwd*. A naive representation of this campaign could consider all techniques as necessary steps (in any order) for the attack campaign to succeed—that is represented by $T_1$ in Fig. 3. A more nuanced representation could require only one of the spearphishing techniques to gain initial access—that is done in $T_2$ via an OR gate as intermediate node. A priori, this ambiguity (whether to use $T_1$ or $T_2$ to model the MITRE data) needs to be resolved by the modeler for each campaign. In Sec. 6 we introduce a method that resolves all ambiguity, producing an overarching AT structure for the MITRE ATT&CK knowledge base, that can be instantiated on every possible (present or future) campaign, further parameterised on the desired attack difficulty.

In the following we introduce notation to define attack trees formally, and we use it to manually model two instances of campaigns recorded in the MITRE knowledge base: Wocao and Dream Job.

## 3.1 Attack tree models

*Definition 3.1.* An *attack tree* (AT) $T$ is a tuple $(N, E, t)$ where $(N, E)$ is a rooted directed acyclic graph, and $t\colon N \longrightarrow \{\mathtt{OR}, \mathtt{AND}, \mathtt{BAS}\}$ is a function such that for $v \in N$, it holds that $t(v) = \mathtt{BAS}$ if and only if $v$ is a leaf.

Moreover, $ch\colon\ N \longrightarrow \mathscr{P}(N)$ gives the set of *children* of a node; if $u \in ch(v)$ we call $u$ a child of $v$, and $v$ a parent of $u$. Furthermore, $T$ has a unique root, denoted $R_T$. We wrote $\mathrm{BAS}_T$ for the set of leaves, i.e., basic attack steps. We will omit the subindex $T$ from $R$, BAS, etc. if no ambiguity arises.We let $v = \mathrm{AND}(v_1, \ldots, v_n)$ if $t(v) = \mathtt{AND}$ and $ch(v) = (v_1, \ldots, v_n)$, and analogously for OR, denoting $ch(v) = \{v_1, \ldots, v_n\}$. We denote the universe of ATs by $\mathscr{T}$ and call $T \in \mathscr{T}$ *tree-structured* if for any two nodes $u$ and $v$ none of their children is shared, else we say that $T$ is *DAG-structured*. The semantics of a AT is defined by its successful attack scenarios, in turn given by its structure function. First, the notion of attack is defined:

*Definition 3.2.* An *attack scenario*, or shortly an *attack*, of a static AT $T$ is a subset of its basic attack steps: $A \subseteq \mathrm{BAS}_T$. We denote by $\mathscr{A}_T = 2^{\mathrm{BAS}_T}$ the universe of attacks of $T$. We omit the subscript when there is no confusion.

The structure function $f_T(v, A)$ indicates whether the attack $A \in \mathscr{A}$ succeeds at node $v \in N$ of $T$. For Booleans we adopt $\mathbb{B} = \{1, 0\}$.

*Definition 3.3.* The *structure function* $f_T \colon N \times \mathscr{A} \to \mathbb{B}$ of a static attack tree $T$ is given by:

$$f_T(v, A) = \begin{cases} 1 & \text{if } t(v) = \mathtt{OR} \quad \text{and } \exists u \in ch(v).f_T(u, A) = 1, \\ 1 & \text{if } t(v) = \mathtt{AND} \quad \text{and } \forall u \in ch(v).f_T(u, A) = 1, \\ 1 & \text{if } t(v) = \mathtt{BAS} \quad \text{and } v \in A, \\ \mathtt{0} & \text{otherwise.} \end{cases}$$

An attack $A$ is said to *reach* a node $v$ if $f_T(v, A) = 1$, i.e. it makes $v$ succeed. If no proper subset of $A$ reaches $v$, then $A$ is a *minimal attack on* $v$. The set of minimal attacks on $v$ is denoted $[\![v]\!]$. We define $f_T(A) \doteq f_T(R_T, A)$, and attacks that reach $R_T$ are called *successful* with respect to $T$. Furthermore, the minimal successful attacks on $R_T$ are called *minimal attacks*. ATs are *coherent* [4], meaning that adding attack steps preserves success: if $A$ is successful then so is $A \cup \{a\}$ for any $a \in \mathrm{BAS}$. This means that the collection of successful attacks of an AT is characterised by its minimal attacks alone.

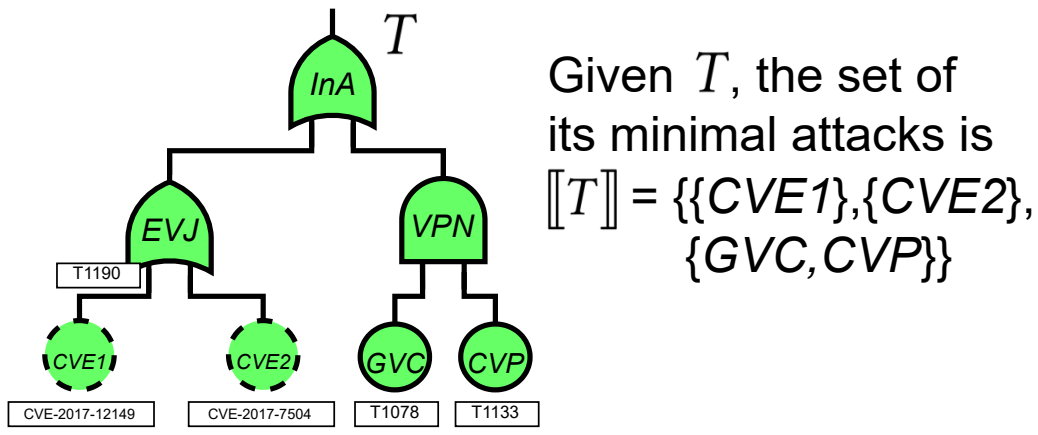

Fig. 4. Minimal attacks of the AT for the *Initial Access* tactic of MITRE's Wocao campaign (excerpt from Fig. 5).

*Definition 3.4.* The *semantics of an AT* $T$ is its collection of minimal attacks $[\![T]\!]$.

*Example 3.5.* Consider the AT in Fig. 4 representing initial access as recorded in MITRE's Wocao campaign (excerpt of Fig. 5): its collection of minimal attacks is $\big\{\{CVE1\}, \{CVE2\}, \{GVC, CVP\}\big\}$. That is, to mount a minimal attack, a malicious actor needs to gain access either by exploiting vulnerabilities in JBoss *EVJ* — via CVE-2017-12149 or CVE-2017-7504 — or by getting valid credentials *GVC* and connecting to the network via VPN (the *CVP* BAS).

*AT-based threat model.* The set of minimal attacks helps *defenders* in cybersecurity risk management: minimal attacks consisting of few BASes represent main security threats, and systems with many possible minimal attacks may be especially vulnerable. Note that, to get a complete picture of the system's vulnerability, this needs to be combined with expert knowledge about the likelihood of each BAS. A quantitative approach to this is given by AT metrics, which we discuss in Sec. 4.

Moreover, an AT considers *attacker actions* (only) by representing the system's attack state. Defender actions are passive in these models, i.e. incorporated in the AT by requiring attackers to perform multiple steps to circumvent in-place preventive measures (AND-gates). For *dynamic defender actions*, modelling formalisms such as *attack–defense trees* have been proposed [18].

While MITRE does propose countermeasures for several attack techniques, most are system-specific and many are actually passive, e.g. configuration files at system deploy-time such as code-signing. In part this is because reactive defenses can be less effective, and require expensive resources such as Security Operations Centers. We thus model attacker actions explicitly, and consider a passive role of the defender, who can design and analyse the system (before deployment): for this, AT models are specially well suited.

## 3.2 Operation Wocao

In this section, we introduce the first *ad hoc* AT model manually built from MITRE ATT&CK. It represents the Wocao campaign[‡]: a cyber-espionage campaign targeting organizations around the world—including Brazil, China, France, Germany, Italy, Mexico, Spain, the United Kingdom, and the United States [29, 46]. The suspected China-based actors have compromised government organizations and managed service providers, as well as aviation, construction, energy, finance, health care, insurance, offshore engineering, software development, and transportation companies.

[‡] Named after a command-line entry typed by one of the threat actors, possibly out of frustration from losing shell access to the target — 我操 *wǒ cào* is Mandarin for 'damn' — see https://attack.mitre.org/versions/v14/campaigns/C0014/.

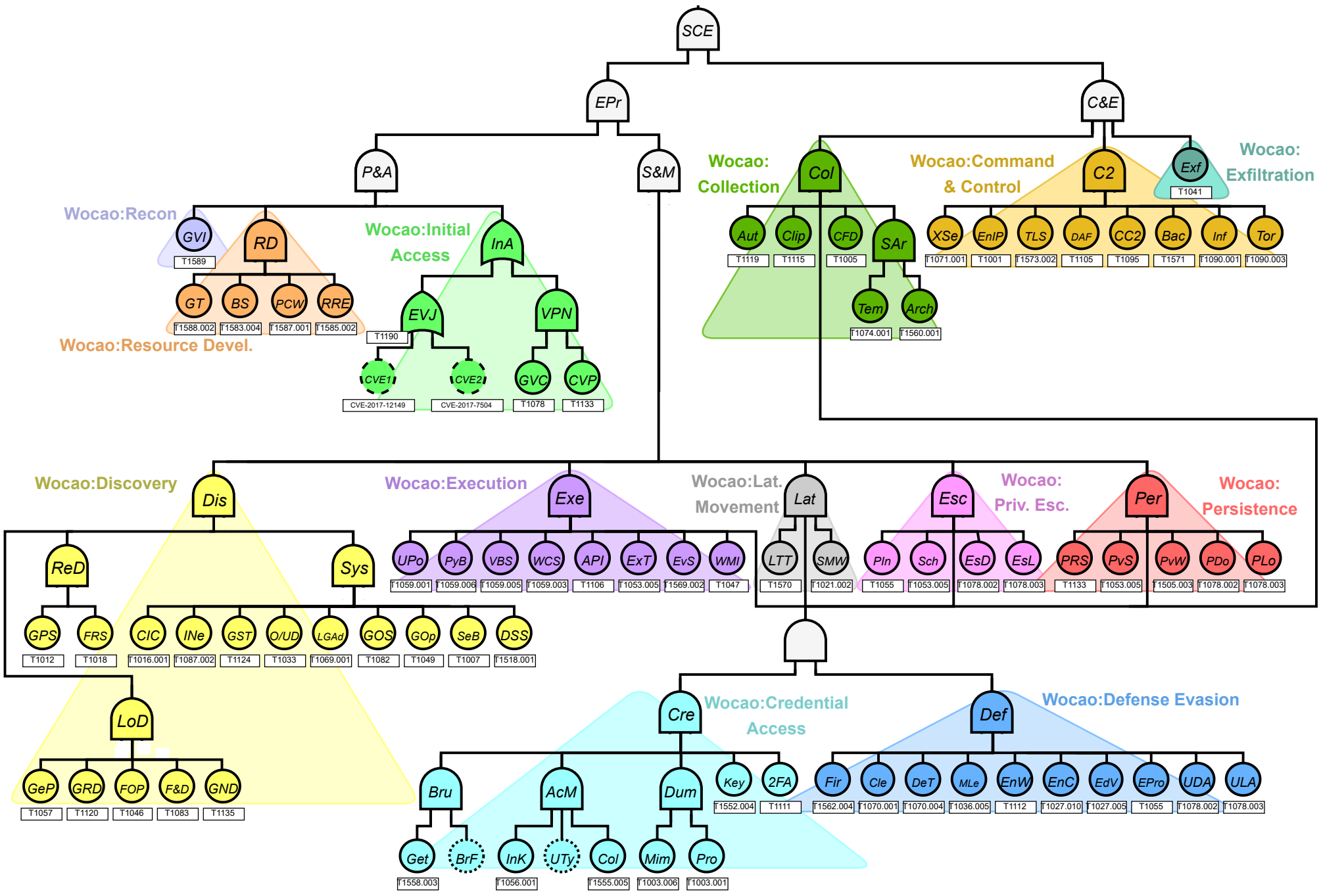

Fig. 5. AT custom model for the Wocao campaign from MITRE [https://attack.mitre.org/versions/v14/campaigns/C0014/]. The meaning of the abbreviations in the figure can be found in Table 1 on page 12.

*Specific modelling choices.* As hinted, MITRE includes technical comments and external references that ground campaigns records: these provide (possibly unstructured) empirical information on some of the steps carried out in a specific campaign. Thus, we ground our *ad hoc* model of the Wocao campaign in the publicly available technical report at [46]. The report presents Wocao as a cyber-espionage operation, where the end goal is to collect and exfiltrate sensitive information. Therefore, the tactical goals *Defense Evasion* and *Credential Access*—which are the basis of the end objective—ground almost every other tactic, including *Lateral Movement* and *Collection*. This was modelled as a DAG structure, placing both ground tactics inside of a module [9], i.e. they have a single (AND gate) parent that connects them to the rest of the AT.

Another point of decision concerns the *Initial Access* tactic, which by default includes technique *Valid Accounts* (`T1078`), as well as its sub-techniques *Domain Accounts* (`T1078.002`) and *Local Accounts* (`T1078.003`). However, a technical comment states that these credentials "found during intrusion" were used for "lateral movement and privilege escalation", and the MITRE tactics *Lateral Movement* and *Privilege Escalation* do list the sub-techniques as part of their steps. As a result, for tactic *Initial Access* only the parent technique `T1078` was kept, because MITRE does not mention which specific sub-technique is needed for initial access, but obtaining (any) valid VPN credential is a known requirement [46].

Finally, the report includes a natural language description of two steps, brute-forcing a password or having a user typing in the master password, that can result in credential access spills [46]. These were modelled as two BASes under tactic *Credential Access*, to showcase an exemplar situation of AT nodes that require expert-guided likelihood attribution depending on the specific setting at hand. Similarly, if more precise information is known (e.g. the CVE IDs that correspond to a specific technique), then data-based probability values can be used, such as their EPSS value.

A complete and detailed description of every modelling decision made in creating the custom AT for the Wocao campaign is included in Appendix A.

*Resulting AT model.* Fig. 5 depicts the AT resulting from our modelling of the Wocao campaign from MITRE, as per the rules and expert-decisions listed above.

Table 1. Abbreviations for the AT in Fig. 5.

| **Wocao Campaign TLE:** | | Discover Security Software | DSS | Collect Credentials | Col |
|---|---|---|---|---|---|
| Success in Cyber Espionage | SCE | Get OS Conn. Systems | GOS | Dump Creds from Memory | Dum |
| Establish Presence | EPr | Get Open Connections | GOp | Dump Creds with Mimikatz | Mim |
| Collect & Exfiltrate | C&E | Search Backdoors | SeB | Dump Creds with ProcDump | Pro |
| Prep. & Access | P&A | **Execution:** | | Get Certificates and Priv. Keys | Key |
| Spread & Maintain | S&M | Execution | Exe | Intercept 2FA Tokens | 2FA |
| **Reconnaissance:** | | Use Powershell | UPo | **Defense Evasion:** | |
| Gather Victim Info | GVI | Create Python Backdoors | PyB | Defense Evasion | Def |
| **Resource Devel.:** | | VBScript Recon | VBS | Modify System Firewall | Fir |
| Resource Devel. | RD | Windows Command Shell | WCS | Clean Logs | Cle |
| Get Tools | GT | Inject with Native API | API | Delete Traces | DeT |
| Buy Server with BTC | BS | Exec with Scheduled Tasks | ExT | Match Legit Names | MLe |
| Prepare Custom Webshell | PCW | Exec via Services | EvS | Enable Wdigest via Registry | EnW |
| Register Rogue Email Addr. | RRE | Execute via WMI | WMI | Encode/Compress Powershell | EnC |
| **Initial Access:** | | **Lateral Movement:** | | Edit Var Names with Impacket | EdV |
| Initial Access | InA | Lateral Movement | Lat | Evade via Proc. Injection | EPro |
| Exploit Vulns. in JBoss | EVJ | Lateral Tool Transfer | LTT | Use Valid Domain Accounts | UDA |
| Exploit CVE-2017-12149 | CVE1 | Use SMB/Win Admin Shares | SMW | Use Valid Local Accounts | ULA |
| Exploit CVE-2017-7504 | CVE2 | **Privilege Escalation:** | | **Collection:** | |
| VPN in Network | VPN | Privilege Escalation | Esc | Collection | Coll |
| Get Valid Creds | GVC | Process Injection | PIn | Auto. Collection via Script | Aut |
| Connect to VPN | CVP | Escalate via Scheduled Tasks | Sch | Collect Clipboard Data | Clip |
| **Discovery:** | | Escalate via Domain Accounts | EsD | Collect Files & Dirs | CFD |
| Discovery | Dis | Escalate via Local Accounts | EsL | Stage and Archive Data | SAr |
| Local Discovery | LoD | **Persistence:** | | Stage Data in Temp Dir | Tem |
| Remote Discovery | ReD | Persistence | Per | Archive Data with WinRAR | Arch |
| Get System Info | Sys | Persist via Remote Services | PRS | **Command & Control:** | |
| Get Processes | GeP | Persist via Scheduler | PvS | Command & Control | C2 |
| Get Removable Disks | GRD | Persist via Web Shell | PvW | Communicate via XServer | XSe |
| Find Open Ports | FOP | Persist via Domain Accounts | PDo | Encrypt IP Addresses | EnIP |
| Files & Dirs Scan | F&D | Persist via Local Accounts | PLo | Upgrade to TLS Socket | TLS |
| Get Network Disks | GND | **Credential Access:** | | Download Additional Files | DAF |
| Get PuTTY Sessions | GPS | Credential Access | Cre | Use Custom Protocol for C2 | CC2 |
| Find Remote Systems & Subnets | FRS | Bruteforce Passwords | Bru | Use High Ports for Backdoor | Bac |
| Check Internet Conn. | CIC | Get Passwd with PowerSploit | Get | Route via Infected Systems | Inf |
| Info with 'net' | INe | Bruteforce Offline | BrF | Route via Tor | Tor |
| Get System Time | GST | Access Passwd Manager | AcM | **Exfiltration:** | |
| Owner/User Discovery | O/UD | Install Keylogger | InK | Exfil. Over C2 Channel | Exf |
| Get Local Group Admins | LGAd | User Types Master Passwd | UTy | | |

### 3.3 Operation Dream Job

Let us introduce a second *ad hoc* AT model for Operation Dream Job: a cyber espionage operation likely conducted by the *Lazarus Group* that targeted the defense, aerospace, government, and other sectors in the United States, Israel, Australia, Russia, and India https://attack.mitre.org/versions/v14/campaigns/C0022/. This model was introduced in Nicoletti et al. [36].

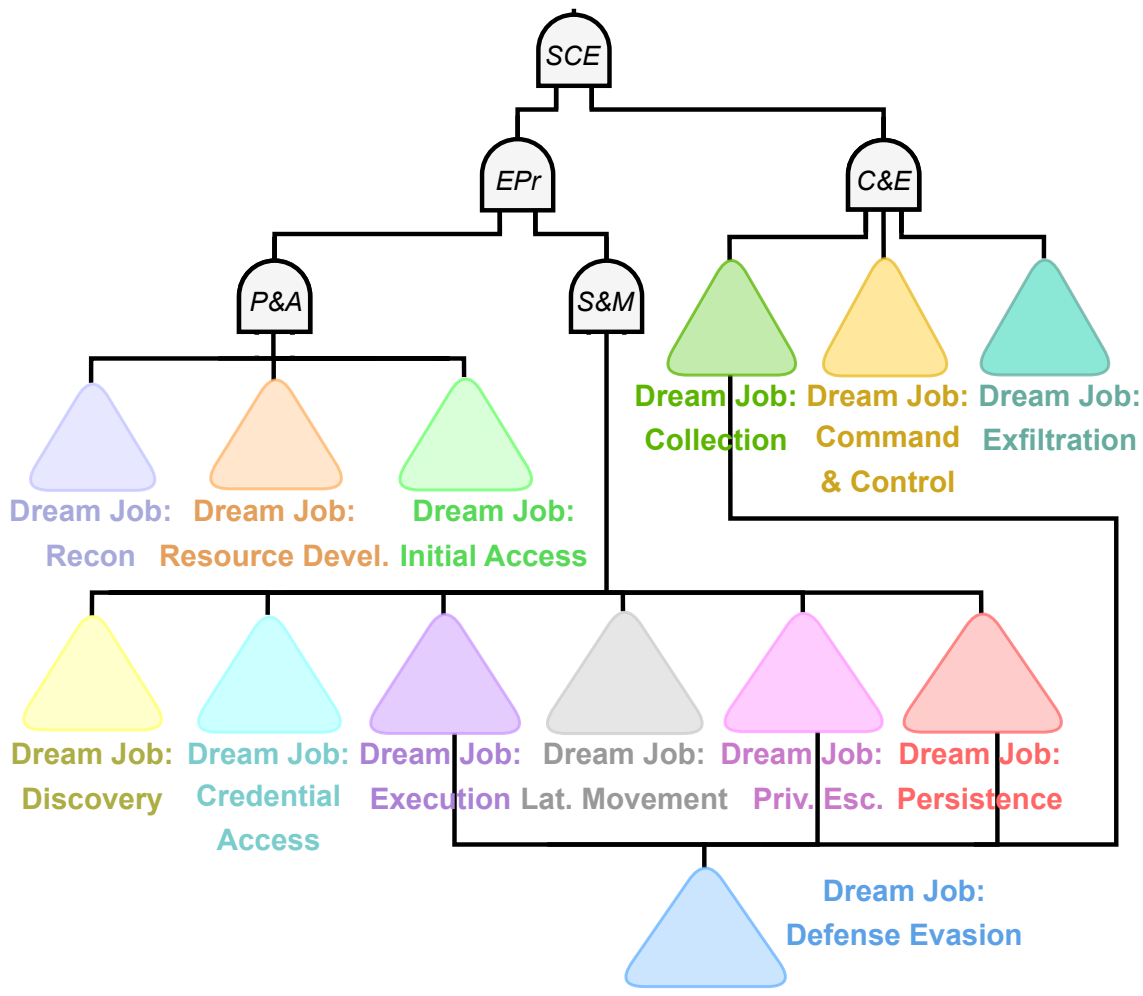

Fig. 6. Skeleton for the AT custom model for the Dream Job MITRE campaign [https://attack.mitre.org/versions/v14/campaigns/C0022/]. For the full AT model see Fig. 13 in Appendix B.

*Specific modelling choices.* Nicoletti et al. [36] treat every (sub)technique in the ATT&CK Navigator as a BAS: in Fig. 13, a small label for each BAS signals its respective (sub)technique according to the MITRE nomenclature. Furthermore, they assume that representing what we define as more fine-grained information (see Sec. 2.2.3) will be more useful than selecting the coarse-grained equivalent. E.g., if the ATT&CK Navigator for tactic *Reconnaissance* presents both the *Gather Victim Org Information* technique (`T1591`) and its subtechnique *Identify Roles* (`T1591.004`) authors keep the latter as an AT node and discard the former, due to *Identify Roles* (`T1591.004`) giving more information on the specifics of this attack step. Finally, authors color code nodes (see legenda at page 35) in the AT following MITRE's 14 tactics (MITRE does not report any technique affiliated to the *Impact* tactic for this campaign). We follow the same colouring convention in our *ad hoc* AT model for Wocao. A detailed representation of the AT can be found in Appendix B.

*Resulting AT model.* Fig. 6 shows a schematic representation of the AT introduced in [36]. The detailed AT model can be found in Appendix B.

## 4 SECURITY METRICS FOR ATTACK TREES

Security metrics — such as the minimal time and cost among all attacks — are essential to perform quantitative analysis of systems and to support more informed decision making processes. For instance, the need of a company to employ a security operations center (SOC) could be defined by legal regulations, e.g. in the case of government bodies, but also by the expected likelihood to suffer cybersecurity attacks. Computing the probability of successful cyberattacks to the IT infrastructure provides a fact-based quantification of such value. Similarly, the minimum time required to execute one such attack is a relevant metric for SOC operations. For MITRE ATs, statistics on the techniques observed across campaigns yield probabilities – as computed in Sec. 2.2 – that can be used as attributes of the BASes in an AT.

### 4.1 Attack tree metrics

To compute security metrics we adopt the well-established *semiring* framework. Semirings have vast applicability potential [11] and have been successfully used to construct attribute domains on ATs [6, 13, 19, 26]. In this paper – coherently with [34] – we adopt *linearly ordered unital semiring attribute domains* (LOADs) where $V$ is the value domain, $\vartriangle$ is an operator to combine values of BASes in an attack, $\triangledown$ is an operator to combine values of different attacks and $\preceq$ is an order to compare values. LOADs provide a convenient way to define an ample class of metrics including "min cost", "min time" — both with parallel or sequential attack steps — "min skill" and "discrete probability".

*Definition 4.1.* A *linearly ordered unital semiring attribute domain* (simply *attribute domain* or *LOAD*) is a tuple $L = (V, \triangledown, \vartriangle, 1_\triangledown, 1_\vartriangle, \preceq)$ where:

- $V$ is a set;
- $\triangledown, \vartriangle : V^2 \to V$ are commutative, associative binary operations on $V$;
- $\vartriangle$ distributes over $\triangledown$, i.e., $x \vartriangle (y \triangledown z) = (x \triangledown y) \vartriangle (x \triangledown z)$ for all $x, y, z \in V$;
- $\triangledown$ is absorbing w.r.t. $\vartriangle$, i.e., $x \triangledown (x \vartriangle y) = x$ for all $x, y \in V$;
- $1_\triangledown$ and $1_\vartriangle$ are unital elements, i.e., $1_\triangledown \triangledown x = 1_\vartriangle \vartriangle x = x$ for all $x \in V$;
- $\preceq$ is a linear order on $V$.

As anticipated, many relevant metrics for security analyses on ATs can be formulated as attribute domains. Table 2 shows examples, where $\mathbb{N}_\infty = \mathbb{N} \cup \{\infty\}$ includes 0 and $\infty$.

Table 2. AT metrics with attribute domains.

| Metric | $V$ | $\triangledown$ | $\vartriangle$ | $1_\triangledown$ | $1_\vartriangle$ | $\preceq$ |
|---|---|---|---|---|---|---|
| min cost | $\mathbb{N}_\infty$ | min | $+$ | $\infty$ | 0 | $\leq$ |
| min time (sequential) | $\mathbb{N}_\infty$ | min | $+$ | $\infty$ | 0 | $\leq$ |
| min time (parallel) | $\mathbb{N}_\infty$ | min | max | $\infty$ | 0 | $\leq$ |
| min skill | $\mathbb{N}_\infty$ | min | max | $\infty$ | 0 | $\leq$ |
| max prob. | $[0, 1]$ | max | $\cdot$ | 0 | 1 | $\leq$ |

*Example 4.2.* An example of a LOAD is $(\mathbb{N}_\infty, \min, +, \infty, 0, \leq)$. Indeed, min and $+$ are commutative, associative operations on $\mathbb{N}_\infty$. The distributive property amounts to the fact that $x + \min(y, z) = \min(x + y, x + z)$, while the absorbing property can be stated as $\min(x, x + y) = x$. The units are given by $1_{\min} = \infty$ and $1_+ = 0$, and $\leq$ is a linear order on $\mathbb{N}_\infty$. As we will discuss in Ex. 4.4, this LOAD corresponds to the *min cost* metric on ATs.

Note that, while derived metrics such as stochastic analyses and Pareto frontiers are semirings, they are not LOADs [26]. Moreover, some meaningful metrics, such as the cost to defend against all attacks, are not even semirings [6, 27].

To render this framework functional, all BASes of ATs are enriched with attributes. More precisely, first an *attribution* $\alpha$ assigns a value to each BAS; then a *security metric* $\widehat{\alpha}$ assigns a value to each attack scenario; and finally the *metric* $\check{\alpha}$ assigns a value to the set of minimal attacks. We then refer to LOADs to define AT metrics. Given a LOAD $(V, \triangledown, \vartriangle, 1_\triangledown, 1_\vartriangle, \preceq)$ we assign to each BAS $a$ an *attribute value* $\alpha(a) \in V$. The operators $\triangledown, \vartriangle$ are then used to define a metric value for $T$ as follows:

*Definition 4.3.* Let $T$ be an AT and let $L = (V, \triangledown, \vartriangle, 1_\triangledown, 1_\vartriangle, \preceq)$ be a LOAD.

(1) An *attribution on $T$ with values in $L$* is a map $\alpha \colon \mathrm{BAS}_T \to V$;

(2) Given such $\alpha$, define the *metric value* of an attack $A$ by

$$\widehat{\alpha}(A) = \bigtriangleup_{a \in A} \alpha(a);$$

(3) Given such $\alpha$, define the *metric value* of $T$ by

$$\check{\alpha}(T) = \bigtriangledown_{A \in [\![T]\!]} \widehat{\alpha}(A) = \bigtriangledown_{A \in [\![T]\!]} \bigtriangleup_{a \in A} \alpha(a).$$

*Example 4.4.* Consider $L$ from Ex. 4.2 representing the metric *min cost*, and let $T$ be the AT in Fig. 7. We give a cost to each BAS, with the attribution $\alpha = \{\mathit{CVE1} \mapsto 8, \mathit{CVE2} \mapsto 3, \mathit{GVC} \mapsto 11, \mathit{CVP} \mapsto 1\}$. As in Ex. 3.5, $T$ has three minimal attacks: $A_1 = \{\mathit{CVE1}\}$, $A_2 = \{\mathit{CVE2}\}$ and $A_3 = \{\mathit{GVC}, \mathit{CVP}\}$. Since $\vartriangle = +$, we have $\widehat{\alpha}(A_3) = \alpha(\mathit{GVC}) + \alpha(\mathit{CVP}) = 11 + 1 = 12$; this is the cost an attacker needs to spend to perform attack $A_3$. We then calculate $\breve{\alpha}(T) = \min(\widehat{\alpha}(A_1), \widehat{\alpha}(A_2), \widehat{\alpha}(A_3)) = 3$. Indeed, the minimal cost incurred by an attacker to successfully attack the system is by performing the cheapest minimal attack, which is $A_2$.

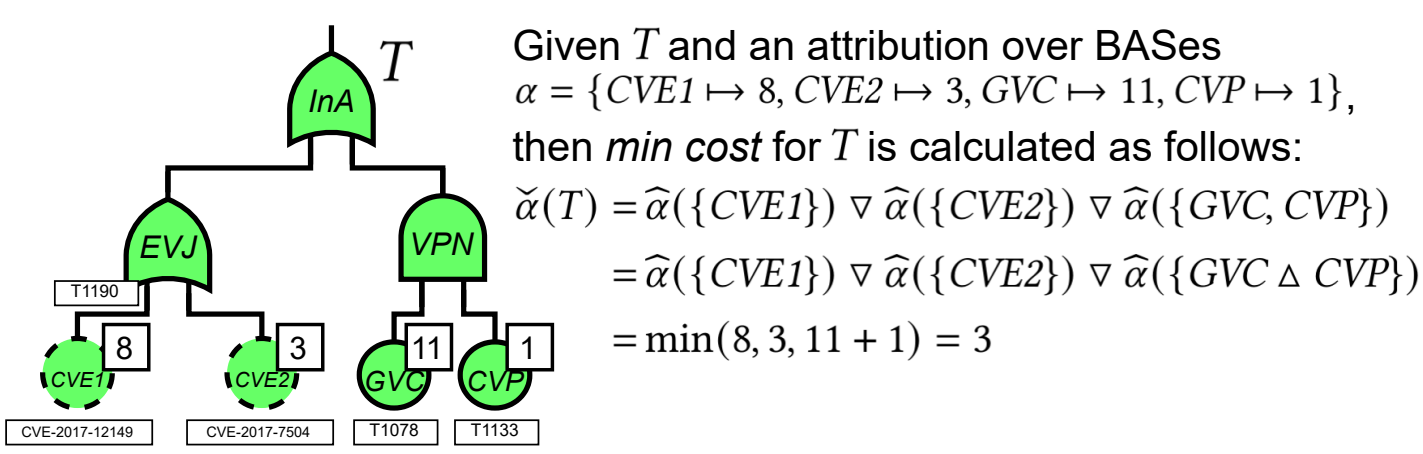

Fig. 7. *Min cost* computation on AT for the `Initial access` tactic of MITRE's Wocao campaign (excerpt from Fig. 5).

When computing multiple metrics on a given AT, one can resort to multiple LOADs and coherently chosen attributions over its BASes. We thus define such a tree as follows:

*Definition 4.5.* An *attributed AT* is a tuple $\mathsf{T} = (T, \mathscr{L}, \mathfrak{a})$ where $T$ is an attack tree, $\mathscr{L} = \{L_1, \ldots, L_l\}$ is a set of LOADs, and $\mathfrak{a} = \{\alpha_i\}_{i=1}^{l}$ is a set of attributions on $T$ s.t. each $\alpha_i$ takes values in $L_i$.

Although in this paper we calculate metrics by considering all *minimal* attacks (see [6, 26]), one could also simply consider all successful attacks. For metrics obtained from LOADs this does not make a difference: for example, the successful attack with minimal cost will always be a minimal attack, since adding BASes can only increase the cost. Therefore, in the calculation of min cost we may as well take the minimum over all successful attacks, rather than just minimal attacks.

## 4.2 Max probability metric

In this paper, we are concerned with probability-based metrics. Specifically, we consider *max probability*, corresponding to the LOAD $([0, 1], \max, \cdot, 0, 1, \leq)$. In other words, given probabilities $\alpha(a) \in [0, 1]$ for every BAS $a$ of an attack tree $T$, the max probability metric of $T$ is defined as

$$\breve{\alpha}(T) \doteq \max_{A \in [\![T]\!]} \prod_{a \in A} \alpha(a).$$

The interpretation is that each BAS has a success probability $\alpha(a)$, and the attacker, aware of the attack tree, attempts the attack that is most likely to succeed, which has success probability $\breve{\alpha}_{\mathrm{mp}}(T)$.

The *max probability* metric expresses the vulnerability of the system in terms of an omniscient attacker, who knows the AT and BASes probabilities, and can thus attack with maximal efficiency. This setting, that is inherent to the metric structure and mirrors the information in MITRE, corresponds to a worst-case scenario that provides a conservative security analysis based on a safe upper bound of (known) attacker capabilities.

## 4.3 Interval arithmetic and p-boxes

In practice, some probabilities $\alpha(a)$ are only approximately known. One way to model this is that only a lower bound $\iota^-(a)$ and an upper bound $\iota^+(a)$ are known, so that $\alpha(a)$ is known to be in the interval $I_a = [\iota^-(a), \iota^+(a)]$; this is also known as a *p-box* [12]. Thus, given $\iota^-$ and $\iota^+$, the set of *feasible attributions* $\mathscr{F}$ is defined as

$$\mathscr{F} = \{\alpha \colon \mathrm{BAS}_T \to V \mid \forall a \colon \iota^-(a) \leq \alpha(a) \leq \iota^+(a)\}.$$

In this case, the set of metric values also forms an interval, given by

$$I_{\mathscr{F}} = \left[ \inf_{\alpha \in \mathscr{F}} \breve{\alpha}(T), \sup_{\alpha \in \mathscr{F}} \breve{\alpha}(T) \right].$$

In principle, finding this interval requires optimization over the multidimensional domain $\mathscr{F}$. However, for max probability, $\breve{\alpha}(T)$ is monotonous in each $\alpha(a)$. This implies that the extrema are obtained when one takes extreme values for each $\alpha(a)$, as is summarized in the following proposition:

Proposition 4.6. *For any LOAD one has* $\inf_{\alpha \in \mathscr{F}} \breve{\alpha}(T) = \widetilde{\iota^-(T)}$ *and* $\sup_{\alpha \in \mathscr{F}} \breve{\alpha}(T) = \widetilde{\iota^+(T)}$.

Thus, in order to calculate $I_{\mathscr{F}}$, we just need to calculate the metric values for $\alpha^-$ and $\alpha^+$. One can do this by interval arithmetic, under which intervals form an attack tree metric itself [26]; this is implemented in Python as *probability bounds analysis* (PBA) or p-box analysis [12]. For completeness, we give the formal definition of an *interval-attributed LOAD*, i.e., the equivalent of definition 4.5 where all attributions are in intervals:

*Definition 4.7.* An *interval-attributed AT* is a tuple $\mathsf{T} = (T, \mathscr{L}, \mathfrak{i})$ where $T$ is an attack tree, $\mathscr{L} = \{L_1, \ldots, L_l\}$ is a set of LOADS, and $\mathfrak{i} = \{\iota_i\}_{i=1}^{l}$ is a set of pairs of attributions $\iota_i = (\iota_i^-, \iota_i^+)$ s.t. each $\iota_i^{\pm}$ takes values in $L_i$, and $\iota_i^-(a) \preceq_i \iota_i^+(a)$ for all $i \leq l$ and $a \in \mathrm{BAS}_T$. For an interval-attributed AT $\mathsf{T}$, we define its $i$-th metric value as

$$\breve{\iota}_i(T) = \left[ \bigtriangledown_{A \in [\![T]\!]} \bigtriangleup_{a \in A} \iota_i^-(a), \bigtriangledown_{A \in [\![T]\!]} \bigtriangleup_{a \in A} \iota_i^+(a) \right], \tag{2}$$

which is an interval in $L_i$ ($\bigtriangleup$,$\bigtriangledown$ refer to the operations of the semiring $L_i$).

### 4.4 Negative log transformation: security index metric

The probability-based metric used in this paper to approximate attack likelihood depends on our knowledge of the probabilities of basic attack steps, either as a fixed probability $\alpha(a) \in [0, 1]$, or an interval $I_\alpha \subseteq [0, 1]$. Such values can be hard to come by in security analyses, as often one only has records of successful attack steps, and not of failed attempts. In this paper, we estimate BAS probabilities as shown, i.e. by counting how often that BAS occurs in the Mitre database – recall Sec. 2.2.2 for details. This does not directly correspond to the probability that this BAS occurs, since we only sample from observed campaigns that have been made public, rather than the complete cyber-criminal behaviour of the world as a whole. Nevertheless, these results are still useful in an ordinal sense: a system whose AT has a *higher max probability* value will be generally *less secure* than a system with a lower value.

We highlight that these metrics will in general not represent the probability of an attack succeeding in the real world—see [25, 42] and Sec. 2.2 in this work for some discussions on this. Practically, to stress this fact during quantitative comparisons among ATs, we do not use their max probabilities directly, but instead operate with their *security index* defined as $\breve{\beta}(T) \doteq -\log(\breve{\alpha}_{\mathrm{mp}}(T)) \in [0, \infty]$. This metric reverses the order to convey a notion of security rather than vulnerability: the higher $\breve{\beta}(T)$, the more secure the system represented by $T$ is.

This can also be interpreted as a semiring transformation: If we take $\beta(a) = -\log(\alpha(a))$, then

$$\breve{\beta}(T) = \min_{A \in [\![T]\!]} \sum_{a \in A} \beta(a).$$

In other words, the security index transforms max probability to the LOAD $([0, \infty], \min, +, \infty, 0, \leq)$, which is also the semiring used for metrics such as *min cost* [26].

## 5 QUANTIFICATION OF MITRE CAMPAIGNS WITH THE CATM LOGIC

In previous sections we computed and operationalized data-driven likelihood values from MITRE via AT models. We now introduce a method to quantify and compare MITRE campaigns via a formal logic and *complete* MITRE AT templates, i.e., able to represent any possible campaign (to be) recorded in MITRE. The logic from [34], ATM, serves as our baseline: it was designed for full expressiveness over AT metrics, e.g. including first-order quantifiers. Cybersecurity practitioners are however mostly concerned with a single question: *how likely/easy/fast can this attack be?* In the following sections (Sec. 5.1 and Sec. 5.2) we present cATM: a cybersecurity-revised fragment of ATM augmented with uncertainty, designed to meet the needs of IT practitioners and to allow for MITRE campaigns modelling and comparison, with minimal effort from part of the practitioners. This is achieved in complement with our MITRE AT templates (presented in Sec. 6.1). We emphasize that in this paper, cATM is just an adaptation of the more general logic ATM for its application to cybersecurity. For a more thorough overview of the logic's expressiveness and computational efficiency, we refer to [34].

### 5.1 Syntax of cATM

We share the objective from [34] of developing a language directly on tree-shaped models with [33, 35]. But while the syntax of ATM in [34] is structured in four layers, here we consider the first two alone, dropping the first-order quantifiers and delegating the computation of quantities (such as probability) to the algorithms for query verification. Also, the first-layer productions of ATM for setting evidence and computing minimal attacks are removed.

The resulting syntax—see eq. (3)—is naturally simpler to deal with by users with no logical-theoretic background, such as IT practitioners, while retaining the metric-querying capabilities of interest, e.g. *how likely is this attack?* Moreover, we extend the second layer with a production to assign an interval attribute to a BAS, capturing the ability to quantify uncertainty.[*]

We call this two-layers logic cATM, and remark its special-purpose design to compute arbitrary metric e.g. via efficient BDD algorithms [26]. What sets cATM apart from a simple metric computation is that *(i)* user queries can assign attribution values to the BASes, omitting an otherwise unavoidable tree-pruning phase (we show practical applications of this in Sec. 6), and *(ii)* it allows to compose attacks via Boolean logic operators, e.g. to query the probability of suffering this *or* that attack.

Formally, given LOADs $\mathscr{L} = \{L_1, \ldots, L_l\} \ni L_k$ and $m \in V_k$, cATM is formed by the following fragment of the syntax from [34], with the added interval production in layer 2 where $L_a, U_a \in V_k$ can define an interval $I_a$ as in Sec. 4.3, or a single value $L_a = U_a = \alpha(a)$:

$$\begin{aligned} &\text{Layer 1:} \quad \phi ::= e \mid \neg\phi \mid \phi \wedge \phi \\ &\text{Layer 2:} \quad \psi ::= \mathbb{V}_k(\phi) \mid \psi\big[a \overset{k}{\mapsto} [L_a, U_a]\big] \end{aligned} \tag{3}$$

As syntactic sugar, we define operators $\vee, \Rightarrow, \Leftrightarrow$ from $\neg, \wedge$ as is customary in propositional logic. Layer 1 in eq. (3) allows to reason about attacks propagation in an AT. Atomic formula $e$ in $\phi$ represents a BAS or an IE in an AT, and can be combined with the standard Boolean binary operators. Layer 2 manipulates metrics: production $\mathbb{V}_k(\phi)$ computes the metric of a given metric on a $\phi$ formula, while production $\psi\big[a \overset{k}{\mapsto} [L_a, U_a]\big]$ sets the attribution of a given basic event $a$ to an arbitrary value $\alpha(a)$ or interval $I_a$.

*Example 5.1.* We continue Ex. 4.4, so there is just one LOAD $L_1 = (\mathbb{N}_\infty, \min, +, \infty, 0, \leq)$.

[*]We do this via unidimensional intervals—extensions e.g. to probability density functions are straightforward.

(1) Suppose we want to know *min cost* for $T$; this would be expressed as $\mathbb{V}_1(\mathit{InA})$. The subscript 1 refers to the (only) LOAD $L_1$.
(2) Suppose that instead of focusing on *InA*, we want to know the minimum cost the attacker needs to succesfully reach both *EVJ* and *VPN* with cost at most 4. This is $\mathbb{V}_1(\mathit{EVJ} \wedge \mathit{VPN})$.
(3) Finally, suppose that we are still interested in the *min cost* for $T$, but from expert input we have learned that the attribution value of *CVE2* is uncertain and could be anywhere in the interval $[2, 10]$. We express this as $(\mathbb{V}_1(\mathit{InA}))\left[\mathit{CVE2} \overset{1}{\mapsto} [2, 10]\right]$. Again, the superscript 1 refers to the fact that we set the metric value of the LOAD $L_1$.

### 5.2 Semantics of cATM

We sketch the semantics of cATM: The complete, formal treatment is given in Appendix C. For layer 1 of cATM, formulae are evaluated on an attack $A$ and on a tree $T$. Atomic formulae $e$ are satisfied by $A$ and $T$ if the structure function in Def. 3.3 returns 1 with $A$ and $e$ as input. This is extended to general layer 1 formulae through the standard interpretation of $\neg$ and $\wedge$ (see **??** This semantics retains the granular reasoning over ATs allowed by [34]. In particular, we can evaluate whether an attack compromises a particular sub-AT without necessarily *reaching* the TLE.

Semantics for layer 2 require the *interval-attributed trees* of definition 4.7. Computing the evaluation of $\mathsf{Val}_{\mathsf{T}}(\mathbb{V}_k(\phi))$ for a general layer 1 formula $\phi$ on an interval-attributed tree $\mathsf{T}$ is the same as computing an interval metric value as in (2), except that we have to range over $[\![\phi]\!]$, the set of all minimal attacks $A$ such that $A, T \models \phi$. The evaluation $\mathsf{Val}_{\mathsf{T}}\left(\psi\left[a \overset{k}{\mapsto} [L_a, U_a]\right]\right)$ is equal to $\mathsf{Val}_{\mathsf{T}'}(\psi)$, where $\mathsf{T}'$ is obtained from $\mathsf{T}$ by setting $\iota_k^-(a) := L_a$ and $\iota_k^+(a) := U_a$; see Appendix C for details.

*Example 5.2.* We continue Ex. 5.1. We take the attributed tree $\mathsf{T} = (T, \{L_1\}, \{\alpha_1\})$, with $T$ and $\iota_1 = \alpha_1 = \alpha$ as in Ex. 4.4, and $L_1$ as in Ex. 5.1.

(1) Since *InA* is the top-level event of $T$, the evaluation $\mathsf{Val}_{\mathsf{T}}(\mathbb{V}_1(\mathit{InA}))$ is simply the metric value $\check{\alpha}(T)$, which we computed in Example 4.4 to be 3.
(2) To find $\mathsf{Val}_{\mathsf{T}}(\mathbb{V}_1(\mathit{EVJ} \vee \mathit{VPN}))$, we first determine $[\![\mathit{EVJ} \vee \mathit{VPN}]\!]$, the set of minimal attacks that ensure both *EVJ* and *VPN* are compromised. These are $A_1 = \{\mathit{CVE1}, \mathit{GVC}, \mathit{CVP}\}$ and $A_2 = \{\mathit{CVE1}, \mathit{GVC}, \mathit{CVP}\}$. These have costs $\widehat{\alpha}(A_1) = 8+11+1 = 20$ and $\widehat{\alpha}(A_2) = 3+11+1 = 15$, so

$$\mathsf{Val}_{\mathsf{T}}(\mathbb{V}_1(\mathit{EVJ} \vee \mathit{VPN})) = \min(\widehat{\alpha}(A_1), \widehat{\alpha}(A_2)) = 15.$$

(3) Finally, we determine the value of $\mathsf{Val}_{\mathsf{T}}\left(\mathbb{V}_1(\mathit{InA})\left[\mathit{CVE2} \overset{1}{\mapsto} [2, 10]\right]\right)$. This is the same as $\mathsf{Val}_{\mathsf{T}'}(\mathbb{V}_1(\mathit{InA}))$, where in $\mathsf{T}' = \mathsf{T}$, except the value of $[\iota^-(\mathit{CVE}), \iota^+(\mathit{CVE})]$ has been changed to the interval $[2, 10]$. For all other minimal attacks, the metric values are unchanged, hence

$$\check{\iota}^-(\mathit{InA}) = \check{\iota}^-(T) = \min(8, 2, 11 + 1) = 2,$$
$$\check{\iota}^+(\mathit{InA}) = \check{\iota}^-(T) = \min(8, 10, 11 + 1) = 8.$$

It follows that $\mathsf{Val}_{\mathsf{T}}\left(\mathbb{V}_1(\mathit{InA})\left[\mathit{CVE2} \overset{1}{\mapsto} [2, 10]\right]\right) = [2, 8]$.

## 6 MITRE ATTACK TREE TEMPLATES

### 6.1 A template to automatically model any MITRE campaign

Besides frequency counting, MITRE's cybersecurity intelligence allows for modelling campaigns as attack trees – as seen in Sec. 3. Here we introduce an automatic approach to generate MITRE AT templates that are *complete*, in the sense that they capture any possible attack campaign as an AT. We implement this for three interpretations of the data in MITRE, which mirror the difficulty an attacker would face when carrying out a campaign:

- **Hard:** for each tactic, every single technique recorded by MITRE (for that campaign) must succeed in order for the campaign to succeed—this is modelled by an AT which only uses AND gates, see Fig. 8a.
- **Easy:** for each tactic, it suffices to successfully carry out at least one technique (among those recorded by MITRE) for the campaign to succeed—this is modelled by an AT which uses a sequential AND gate (SAND) at top level[§], and OR gates everywhere else, see Fig. 8b.
- **Default:** for each tactic, all high-level techniques recorded by MITRE must succeed, but if evidence exists that several subtechniques may have been used to achieve this, the AT template requires at least one of them to succeed—this is modelled by an AT which uses SAND gates at top and tactic levels, and OR gates to refine high-level techniques into their subtechniques, see Fig. 8c.

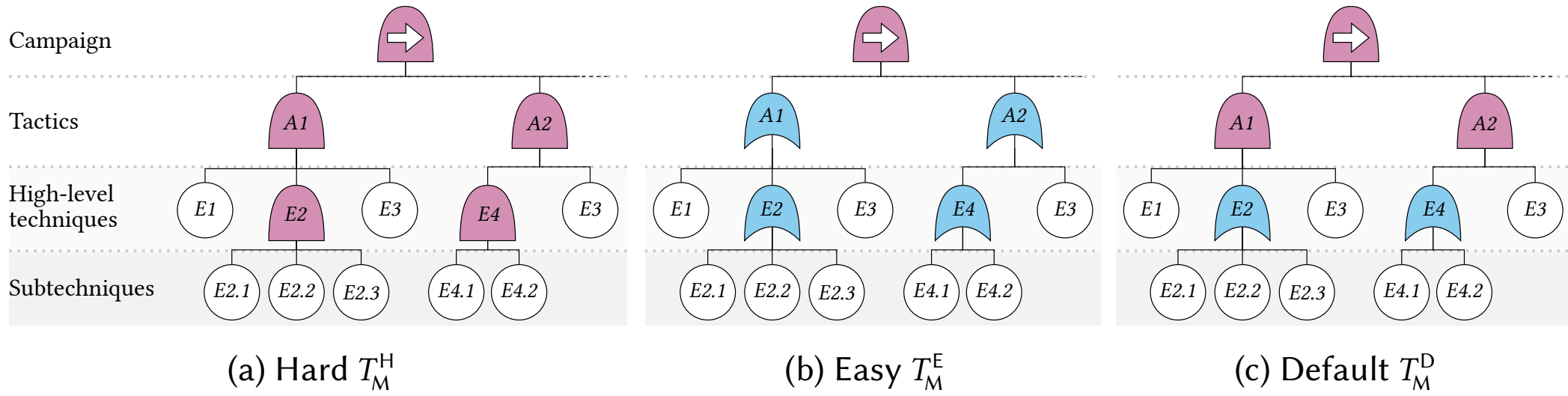

(a) Hard $T_{\mathrm{M}}^{\mathrm{H}}$ (b) Easy $T_{\mathrm{M}}^{\mathrm{E}}$ (c) Default $T_{\mathrm{M}}^{\mathrm{D}}$

Fig. 8. AT templates for MITRE campaigns, in three difficulty levels from the perspective of a threat actor.

Automatic approaches are advantageous w.r.t. crafting ad hoc ATs manually in the usual ways, including being less error-prone and time consuming, easier to update, etc. By using SANDs we assume a total order of the 14 tactics registered in the knowledge base. Then, a generic AT template that can represent any MITRE campaign would consist of the following levels:

**Campaign:** The top-level event—that signifies the success of a campaign—is a SAND gate, whose children are the 14 MITRE tactics;

**Tactics:** Each tactic $A$ is either an AND or an OR gate, whose children are the high-level techniques that MITRE identifies as possible actions to achieve $A$;

**Techniques:** Each technique $E$ is either a BAS (if it is a high-level technique without subtechniques), or a gate whose children are the subtechniques that can be used to deploy $E$.

This yields MITRE AT templates like those in Figs. 8 and 9, that represent any campaign published by MITRE in three possible difficulty levels. Then, to model a campaign $C$ for which specific tactics and (sub)techniches have been recorded, the instantiated AT (e.g. $T_{\mathrm{M}}^{\mathrm{D}}$) is pruned from tactics and techniques not recorded for $C$, producing a custom tree $T_C^{\mathrm{D}}$. Then, the resulting BASes are to be quantified with the relevant attributes—e.g. probabilities as per Sec. 2.2—to enable the computation of specific metrics, such as the security index $\breve{\beta}(T_C^{\mathrm{D}})$ from Sec. 4.4. Alternatively, cATM can be used to query that template without the need for middle steps like AT pruning and BAS decoration–we show this next in Sec. 6.2. The advantage of our approach is its simplicity in comparison to (manually and correctly) pruning an labelling to an AT with 800 nodes, including ca. 700 BASes.

## 6.2 Quantification of MITRE campaigns via cATM

Our logic allows for computing the security index of a MITRE campaign $C$ using the MITRE AT templates. This value (e.g. $\breve{\beta}(T_C^{\mathrm{D}}) \in [0, \infty]$ for the example above) serves to compare the likelihood of observing different campaigns in the wild. More precisely, $\breve{\beta}(T_C^{\mathrm{D}}) > \breve{\beta}(T_{C'}^{\mathrm{D}})$ indicates that campaign

[§]SAND gates behave like AND, but further require that its children fail in left-to-right order.

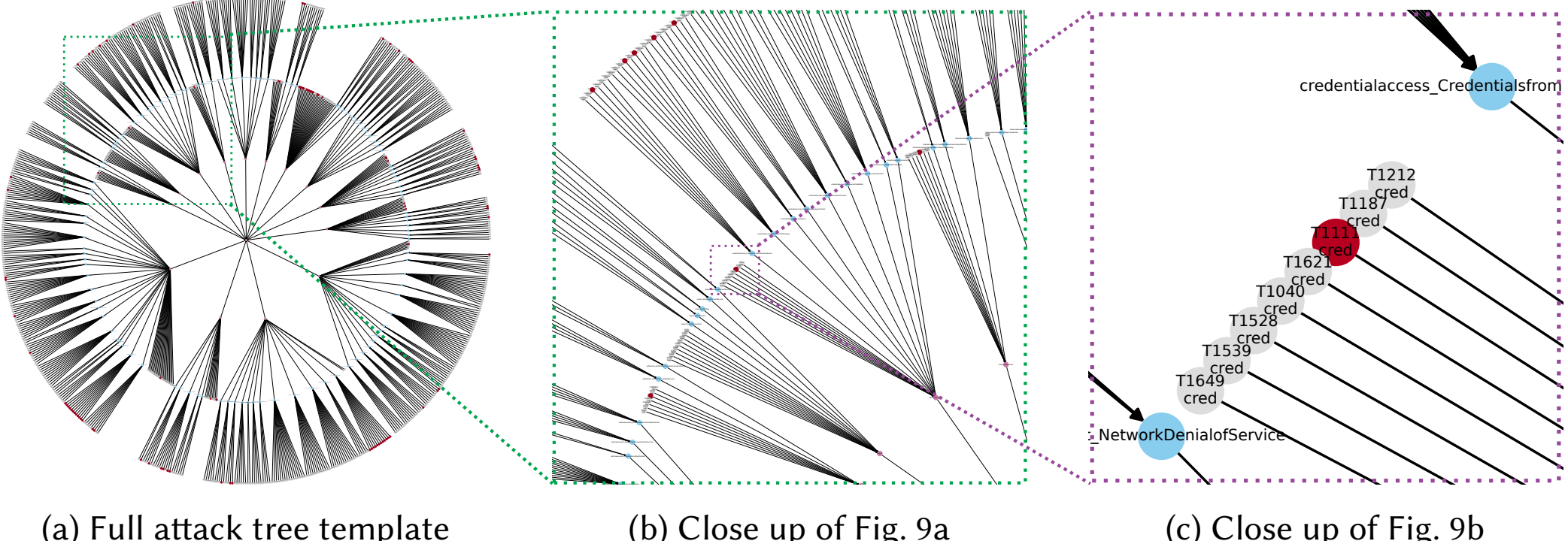

(a) Full attack tree template (b) Close up of Fig. 9a (c) Close up of Fig. 9b

The center of the AT in Fig. 9a is the root node that symbolises a successful campaign. The 14 MITRE tactics are its direct children; high-level techniques are the following level, and subtechniques are the leaves of the graph. Fig. 9c colours dark red the (sub)techniques recorded in MITRE ATT&CK for Wocao.

Fig. 9. MITRE attack tree (default) template for the Wocao campaign

**Input:** Attack tree $T$, node $v$ from $T$, neutral element $1_g$, campaign $\boldsymbol{C}$, probabilities $\boldsymbol{p}$
**Output:** Security index of the campaign $\breve{\beta}(T_C) \in [0, \infty]$

1 **if** $v.\mathsf{t} = \mathsf{AND} \lor v.\mathsf{t} = \mathsf{SAND}$ **then**
2 **return** $\sum_{u \in v.\mathtt{children}} \mathsf{Sec}(T, u, 0, \boldsymbol{C}, \boldsymbol{p})$ // 0 is neutral of AND/SAND for $([0, \infty], \min, +)/\log$
3 **else if** $v.\mathsf{t} = \mathsf{OR}$ **then**
4 **return** $\min_{u \in v.\mathtt{children}} \mathsf{Sec}(T, u, \infty, \boldsymbol{C}, \boldsymbol{p})$ // $\infty$ is neutral of OR for $([0, \infty], \min, +)/\log$
5 **else** // $v.\mathsf{t} = \mathsf{BAS}$
6 **if** $\boldsymbol{C}[v.\mathsf{E}, v.\mathsf{A}] = \bot$ **then** // technique $E$ not used for tactic $A$ in campaign $C$
7 **return** $1_g$ // $1_g$ is the neutral element of the parent gate
8 **else**
9 **return** $-\log\left(\boldsymbol{p}[v.\mathsf{E}, v.\mathsf{A}]\right)$ // $-\log$ of probability of using technique $E$ for tactic $A$

Algorithm 1: Computation of the security index of a MITRE campaign $C$.

$C$ is "more secure" than campaign $C'$, i.e. *less likely* to be exploited by a threat agent due to the techniques required to carry them out. The gist of our approach—that requires no AT manipulations–is to leverage the logic production $\psi[a \overset{k}{\mapsto} [L_a, U_a]]$ to:

*(i)* *turn off* in $T_{\mathrm{M}}$ the techniques which were not used in campaign $C$;
*(ii)* *turn on* and quantify those that were;
*(iii)* compute the metric value at top-level of the AT.

In step *(i)*, for a BAS $a$ corresponding to a technique $E$ not used in $C$, we set $L_a = U_a = \alpha(a)$ to the neutral element of the operator of the parent gate: so 0 for a parent AND-gate (operator: +) and $\infty$ for a parent OR-gate (operator: min). On the other hand, when $E$ is used in $C$, we set $L_a = U_a = \alpha(a) = -\log(p_a)$ in step *(ii)*, where $p_a$ is the probability of technique $E$. In step *(iii)*, we compute the metric value using the logic production $\mathbb{V}_1$; here 1 refers to the fact that we only consider one metric, the security index. We introduce this approach in Algo. 1—which is an adaptation to our setting of the $\mathsf{BU}_{\mathsf{DAT}}$ algorithm from [26]—and illustrate it with the following example.

*Example 6.1.* Assume that the default MITRE AT template from Fig. 8c consists only of the $A_1$ branch, and consider a campaign $C$ where (sub)techniques $E_1$, $E_{2.2}$, and $E_{2.3}$ were used for tactic $A_1$.

To compute the security index of campaign $C$ for the default template, $\breve{\beta}(T_C^{\mathrm{D}})$, we need the probability values of those subtechniques, i.e. $p_{E_1|A_1}, p_{E_{2.2}|A_1}, p_{E_{2.3}|A_1} \in (0,1]$. Algo. 1 keeps this data in a matrix $\boldsymbol{p}$ s.t. $\boldsymbol{p}[E,A] = p_{E|A}$. Similarly, a matrix $\boldsymbol{C}$ represents campaign $C$ s.t. $\boldsymbol{C}[E,A] = \bot$ if technique $E$ was not used for tactic $A$ in that campaign, so in this example we have $\boldsymbol{C}[E_{2.1}, A_1] = \boldsymbol{C}[E_3, A_1] = \bot$. Let $R$ denote the root node of $T_{\mathrm{M}}^{\mathrm{D}}$, i.e. the SAND gate, and note that a single term summation—i.e. line 2 in the algorithm—consists of that term alone. Then computations of Algo. 1 for this example are as follows, where techniques absent from $C$ are assigned the neutral element of the parent gate (i.e. 0 for AND/SAND and $\infty$ for OR, see line 7 in Algo. 1):

$$
\begin{aligned}
&\mathrm{Sec}(T_{\mathrm{M}}^{\mathrm{D}}, A_1, 0, \boldsymbol{C}, \boldsymbol{p}) = \cdots \\
&\quad \cdots = \mathrm{Sec}(T_{\mathrm{M}}^{\mathrm{D}}, E_1, 0, \boldsymbol{C}, \boldsymbol{p}) + \mathrm{Sec}(T_{\mathrm{M}}^{\mathrm{D}}, E_2, 0, \boldsymbol{C}, \boldsymbol{p}) + \mathrm{Sec}(T_{\mathrm{M}}^{\mathrm{D}}, E_3, 0, \boldsymbol{C}, \boldsymbol{p}) \\
&\quad = -\log(\boldsymbol{p}[E_1, A_1]) + \mathrm{Sec}(T_{\mathrm{M}}^{\mathrm{D}}, E_2, 0, \boldsymbol{C}, \boldsymbol{p}) - 0 \\
&\quad = -\log(p_{E_1|A_1}) + \min\left(\mathrm{Sec}(T_{\mathrm{M}}^{\mathrm{D}}, E_{2.1}, \infty, \boldsymbol{C}, \boldsymbol{p}), \mathrm{Sec}(T_{\mathrm{M}}^{\mathrm{D}}, E_{2.2}, \infty, \boldsymbol{C}, \boldsymbol{p}), \mathrm{Sec}(T_{\mathrm{M}}^{\mathrm{D}}, E_{2.3}, \infty, \boldsymbol{C}, \boldsymbol{p})\right) \\
&\quad = -\log(p_{E_1|A_1}) + \min\left(\infty, -\log(\boldsymbol{p}[E_{2.2}, A_1]), -\log(\boldsymbol{p}[E_{2.3}, A_1])\right) \\
&\quad = -\log(p_{E_1|A_1}) + \min\left(\infty, -\log(p_{E_{2.2}|A_1}), -\log(p_{E_{2.3}|A_1})\right) \\
&\quad = -l_{E_1} + \min(-l_{E_{2.2}}, -l_{E_{2.3}}).
\end{aligned}
$$

Note that $l_E = \log(p_{E|A_1}) \in [0, \infty)$ for all $E \in \{E_1, E_{2.2}, E_{2.3}\}$, and thus $\mathrm{Sec}(T_{\mathrm{M}}^{\mathrm{D}}, A_1, 0, \boldsymbol{C}, \boldsymbol{p}) \in \mathbb{R}_{\geqslant 0}$. Also, the two techniques not used in $C$ where assigned values based on their parent gates: while $E_3$ was assigned 0 (the neutral element of +), $E_{2.1}$ was assigned $\infty$ (the neutral element of min).

*Technical considerations of Algo. 1.* The same arithmetic operator (summation) is used to aggregate the result of the children of AND and SAND: this represents that all children must succeed, in any order, for the gate to succeed. In the case of SAND, this represents a conservative upper-bound on the security index value, where tactical goals achieved via orders not contemplated in MITRE can still result in a successful campaign. Requiring specific execution orders would entail differentiating the AND and SAND cases in Algo. 1, and adjusting the semiring framework as discussed in Sec. 4, see also [26]. Probabilities are stored as a matrix $\boldsymbol{p}$ indexable by techniques and tactics s.t. $\boldsymbol{p}[E,A] = p_{E|A}$—this is global data attainable from MITRE public resources, e.g. as described in Sec. 2.2.2. Any campaign $C$ is also encoded as a matrix $\boldsymbol{C}$ s.t. $\boldsymbol{C}[E,A] = \bot$ if $E$ was not used for tactic $A$ in campaign $C$, localising all data about the campaign's techniques in a single object that represents it. Then, any high-level technique $E$ that MITRE refines into subtechniques is represented by a gate, whose type depends on the template chosen (hard, default, or easy). Furthermore, let $E$ be any technique that has no subtechnique, or any subtechnique in MITRE: then $E$ is represented as a BAS $v$ with fields $v.\mathsf{E}, v.\mathsf{A}$ for its corresponding technique and tactic from MITRE ATT&CK. This keeps data localised, concentrating all structural information of MITRE attack representations in the AT $T_{\mathrm{M}}$, to facilitate updates e.g. in case that MITRE changes its attack structure by adding new techniques/tactics. Finally, we highlight that neutral elements 0 and $\infty$ come from the LOAD $([0,\infty], \min, +, \infty, 0, \leq)$, which implicitly encodes the negative log transformation required for our security index computation $\breve{\beta}(\cdot)$—see Sec. 4.4.

## 7 EMPIRICAL DEMONSTRATION ON ALL MITRE ATT&CK CAMPAIGNS

Here we use the theory of the previous sections to compute the security indices of our custom ATs for the Wocao and Dream Job campaigns, as well as their respective modelling via the MITRE AT templates. We then extend this execution to all (23) MITRE ATT&CK Enterprise campaigns. This provides concrete evidence on the performance of our quantification framework, which is easily applicable to a large dataset of cyberattack campaigns, while not numerically comparable to other contributions in the literature—see Sec. 1.1 "Related work" for further discussion on this.

### 7.1 Implementation in software tools

We developed an open-source Python library that offers an abstract data type for attack trees, as well as generic and efficient algorithms for the computation of metrics on AT instances, including a prototypical implementation of the cATM logic that can query the probability of an AT. We make this available in a software reproduction package at DOI `10.5281/zenodo.14193936`. The current implementation is based on data taken from MITRE v14, which was available at the time of writing—this can be substituted for a user-selected version of MITRE.

The `AttackTree` Python library offers the `AttackTree` class, that can be used e.g. to instantiate the custom MITRE campaign ATs for Wocao and Dream Job as introduced in Secs. 3.2 and 3.3. We also offer another Python tool to automatically instantiate any campaign recorded in MITRE according to the template patterns introduced in Sec. 6. To do this, the user passes the campaign code—e.g. `"C0022"` for Dream Job—and the desired difficulty level, which if omitted defaults to `"default"`. Then, computing a metric is done by invoking the proper library function, e.g. `compute_metric("cATM")` for the cATM logic. Code 1 shows how to do this for the Wocao campain, using the default MITRE AT template—specifically line 4 results in the computation introduced in Algo. 1.

```
1 from numpy import log
2 from MITRE_AT_template_creator import MITRE_AT_template
3 AT = MITRE_AT_template("C0014", "default")
4 securityIdx = -log(AT.compute_metric("cATM"))
```

Code 1. Compute security index for MITRE AT (default) template for Wocao

### 7.2 Comparing MITRE campaigns

The MITRE AT templates from Sec. 6 provide a baseline or ground-truth against which to compare possible custom models of a MITRE ATT&CK campaign.

Recall that the hard template represents the hardest possible interpretation of MITRE ATT&CK data grounding each campaign. The probability of a successful attack in this model is very low: it represents a very conservative success probability estimate, and corresponds to a high security index. On the other extreme of the spectrum, the easy template results in a comparatively high probability of successful attack, that corresponds to a low security index. In other words, for a given campaign $C$, a hard template $T_C^{\mathsf{H}}$ obtains a high value $\check{\beta}(T_C^{\mathsf{H}})$, and an easy template obtains a strictly lower value $\check{\beta}(T_C^{\mathsf{E}})$, resulting in a feasible *security range* $\left[\check{\beta}(T_C^{\mathsf{E}}), \check{\beta}(T_C^{\mathsf{H}})\right] \neq \varnothing$ for campaign $C$.

This range provides a baseline for validation and comparison among campaigns: (1) the structure of the MITRE AT template that underlies the model, as well as the probability values with which their leaves are decorated, come directly from MITRE ATT&CK data; (2) this process is automatic and white-box by design—see Secs. 2.2 and 6. Note that the hard template (that uses AND gates only) models a situation where the threat agent needs to carry out every technique recorded in MITRE for that campaign to succeed: this is the strictest possible interpretation of MITRE data, requiring the highest amount of successful techniques. In contrast, the easy template (that uses OR gates) models a situation where the threat agent needs only one (sub)technique per tactic to succeed: this is the laxest interpretation of MITRE data, requiring the lowest amount of techniques to succeed. By the arithmetic operators involved in Algo. 1 this results in the hard template always getting a security index higher than the easy template. As a consequence, the security index of a default template $T_C^{\mathsf{D}}$, that employs a mixture of AND and OR gates, always lies in the security range, i.e. $\check{\beta}(T_C^{\mathsf{D}}) \in \left[\check{\beta}(T_C^{\mathsf{E}}), \check{\beta}(T_C^{\mathsf{H}})\right]$.

*Modelling connection between arbitrary ad hoc ATs and data-driven templates.* Note that all MITRE AT templates from Sec. 6 provide an adherent model of MITRE ATT&CK data, i.e. every recorded technique appears in the AT, every tactic must succeed, and the probability values of the AT leaves are fixed a priori via computations from Mitre data, e.g. as in Sec. 2.2. Conceptually, from a modelling standpoint for a campaign $C$, *the security index of a custom AT model adherent to Mitre ATT&CK data should lie in that range*, because:

(1) the highest security index $\check{\beta}(T_C^{\mathsf{H}})$ requires every (sub)technique recorded for $C$ to succeed, i.e. their probabilities are multiplied, so a higher security index would mean either more techniques (not recorded in Mitre), or different probability values;
(2) analogously, the lowest security index $\check{\beta}(T_C^{\mathsf{E}})$ only requires the (max possible) probabilities of the tactical goals to be multiplied, so a lower security index would either require deeming $C$ successful when only a subset of the tactical goals have been achieved (as opposed to the evidence in Mitre), or different probability values.

Table 3. Comparison of security indices for different models of the Wocao and Dream Job campaigns.

| Campaign | AT model | Security index |
|---|---|---|
| Wocao | Mitre AT template: easy | 24.51 |
| | Mitre AT template: hard | 317.45 |
| | Mitre AT template: default | 184.02 |
| | Custom AT (Sec. 3.2) | 207.61 |
| Dream Job | Mitre AT template: easy | 26.48 |
| | Mitre AT template: hard | 192.10 |
| | Mitre AT template: default | 128.96 |
| | Custom AT (Sec. 3.3) | 180.29 |

*AT models comparison: Wocao.* Table 3 compares the security indices of several AT models—custom models, and Mitre AT templates in their three difficulty levels—for the Wocao and Dream Job campaigns. As expected, the security index of the custom Wocao AT model lies in between the two indices of the corresponding easy and hard templates. Most notably, the default template is the one presenting the closest security index w.r.t. the *ad hoc* AT model from Sec. 3.2. This is due to tactics *Reconnaissance*, *Defence Evasion*, *Privilege Escalation* and *Command and Control*, where the strategy described in Sec. 3 to model a custom AT results in choosing high-level techniques for `T1589`, `T1055` and `T1001`, which is equivalent to the coarse-grained case highlighted in Sec. 2.2.3. In contrast, the AT templates list every subtechnique of these high-level techniques, e.g. `T1055.001` through `.015` for both *Defence Evasion* and *Privilege Escalation*. For the case of the hard template, where high-level techniques are refined via AND gates, this results in a model requiring more subtechniques for the attack to succeed, and thus in a lower probability of attack.

Further differences between that custom AT and the corresponding templates are due to the modelling choice detailed in Sec. 3.2, which uses a DAG structure to put the entire *Credential Access* sub-tree as a child of *Collection*, instead of having the single subtechnique `T1056.001` as direct *Collection* child. This results in an increased probability of attack for the custom AT, since `T1056.001` is represented with a single BAS for both tactics. In contrast, the default template AT requires that technique to succeed at least twice: once for *Credential Access*, and once again for *Collection*: being probability values, their joint occurrence is less likely than any one of them occurring.

The remaining discrepancies in security indices are due to two further custom modelling choices:

- Subtechniques `T1078.002` and `T1078.003` were excluded from tactic *Initial Access*, keeping instead its parent technique *Valid Accounts*. For max probability (Sec. 4.2), this makes the value of the T1078 branch in the *ad hoc* AT to be an upper bound w.r.t. any template AT. In particular, the max probability of that branch in the custom AT is strictly higher than in the hard template.

- The custom choices of adding two BASes for *Credential Access* (*BrF* with probability $p = 0.35$, and *UTy* with $p = 0.85$), and refining T1190 from *Initial Access* ($p = 0.14$ extracted from MITRE), with two specific CVEs from the literature and their EPSS values ($p = 0.97$ and 0.27, of which the max is taken), also result in a higher probability of attack for the custom AT.

*AT models comparison: Dream Job.* Table 3 shows that the security indices of the default MITRE AT template and the custom Dream Job AT from Sec. 3.3 lie, as expected, in the security range given by the easy and hard templates. Most notably, the value for the custom AT is quite close to that of the hard template, but not exactly the same. Inspecting the custom model (Appendix B) we see that the AT includes every technique listed in MITRE for the campaign, but in particular for the *Credential Access* and *Reconnaissance* tactics a high-level technique was used, respectively T1110 and T1589. This is equivalent to the coarse-grained case of data counting from Sec. 2.2.3, which results in multiplying the probability of the most likely subtechnique. In contrast, the hard template lists every subtechnique of those high-level techniques, e.g. T1110.001 through .004 for *Credential Access*, which results in a model that requires more subtechniques to succeed than the custom AT, yielding the aforementioned values. In summary, the hand-built AT is approximately equivalent to the automatically-computed hard template, with the exception of assuming more coarse-grained data in the two aforementioned cases.

*Wocao vs. Dream Job.* Our framework also allows for comparison of security indices between different campaigns. If we consider the Wocao and Dream Job custom models, we can see that Wocao's security index is higher than the security index of Dream Job, representing a greater difficulty – and lesser probability – of successfully carrying out the former campaign w.r.t. the latter. When considering the MITRE AT templates for both campaigns, we see that this order is maintained when comparing the default and hard templates, i.e. the security index of Wocao remains consistently higher than that of Dream Job.

In contrast to the above, for the easy template the security index of Wocao is lower than that of Dream Job, although by an order of magnitude lower than the differences where the values for Wocao are higher. We show next in Fig. 10 that this occurs for many MITRE ATT&CK campaigns, i.e. there is no coherent monotonicity among the security indices of the different-difficulty templates. For instance, arranging in ascending order by security index of default templates does not yield an ascending order of the respective indices for the hard templates. This is due to the hard template using AND gates at the subtechnique level of the ATs, where the default template uses OR gates (Figs. 8a and 8c). Algo. 1 multiplies the subtechniques probabilities in the case of the hard template, and returns the max in the case of the default template. Since the values are probabilities, where max exhibits an absorbing behaviour, multiplication is monotonically decreasing, thus explaining the observed differences.

*Application to all MITRE ATT&CK campaigns.* We can apply the same procedure to every Enterprise campaign in MITRE, which yields the values shown in Fig. 10. The figure arranges campaigns in ascending order of their security indices, which depends on the template used: the left figure sorts the campaigns using the indices computed via the hard templates, i.e. the upper bound of the feasible range that is depicted as the shaded regions; the center figure uses the default templates to sort the campaigns; and the right one uses the easy templates.

These results allow for a straightforward visual comparison of the security indices of all Enterprise campaigns in MITRE ATT&CK. This is a direct indicator of the security of the system under attack, and can be used by SOC analysts, companies, etc. to prioritise resource allocation and defence mechanisms. Moreover, practitioners have all the advantages of a model constructed from a white-box data-driven procedure, including model explainability and customisability whenever desired.

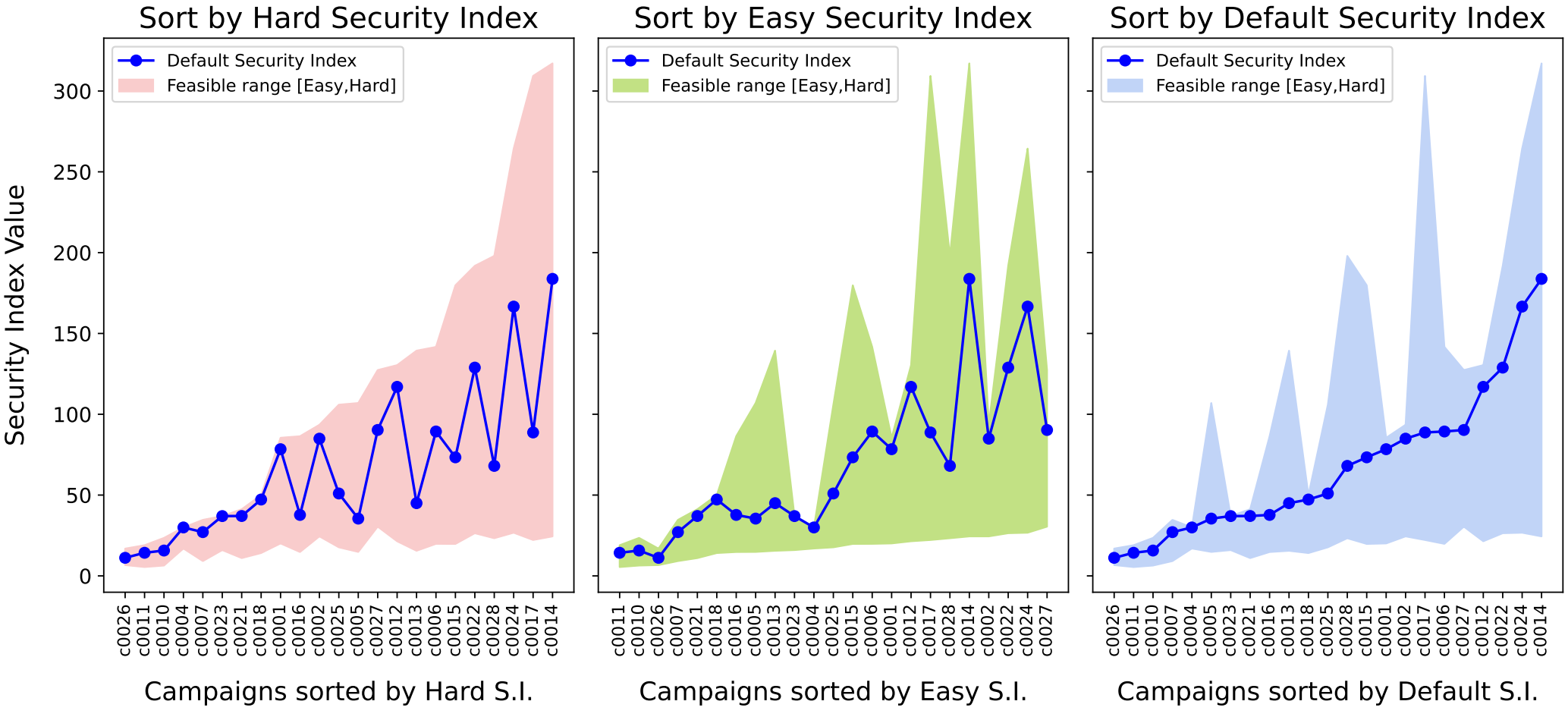

Fig. 10. Default (blue line), easy (lower range), and hard (upper range) values of MITRE Enterprise campaigns

*Comparative baseline.* To compute the security index of a campaign, our framework considers the structure of the recorded attacks using formal tools such as Defs. 4.1 and 4.7. As noted in Sec. 1.1, this is not the only manner in which MITRE ATT&CK campaigns can be quantified. In particular, Kim et al. [16] also use the MITRE Matrix and Lockheed Martin Cyber Kill-Chain, assigning to each tactic a score equal to the number of techniques used in it. This quantification approach is available in the MITRE ATT&CK Navigator, and is insensitive to data granularity nuances such as the ones discussed in Sec. 2.2.3.

To compare this to the approach introduced in Sec. 2.2, take tactic *Initial Access* in Wocao, where the high-level technique *Valid Accounts* (`T1078`) coexists with subtechniques *Domain Accounts* (`T1078.002`) and *Local Accounts* (`T1078.003`). When activating the "aggregate score: sum" in MITRE ATT&CK we get the value 3 for this data, despite the fact that both `T1078.002` and `T1078.003` are special cases of the *Valid Accounts* technique. In general, the aggregate score computed for a campaign in this manner is a linear function of the number of techniques registered for it.

In contrast, while the security index also increases as a function of the number of techniques recorded for a campaign, it is able to distinguish mixed-data and avoid double counting. Thus, the high-level technique *Valid Accounts* is recognised as a generalisation of is two special cases: `T1078.002` and `T1078.003`, whose respective security indices are 2.037 and 2.730. Those two finer-grained values can then be aggregated in different manners, e.g. taking the max or the sum, or any other operation from the LOAD of interest.

Moreover, the same (sub)techniques mentioned above appear for other tactics, e.g. *Privilege Escalation*, where the aggregate score in MITRE ATT&CK again adds +3 to the general total. This counts techniques equally regardless of the tactic for which they are used. Eq. (1) instead distinguishes among tactics, making the probabilities dependent on the frequency with which each technique has been successfully applied to each tactic. In this example, subtechniques `T1078.002` and `T1078.003` respectively get security indices 2.813 and 3.507 for the *Privilege Escalation* tactic, where they have been used with more frequent (recorded) success than for *Initial Access*.

### 7.3 Internal generalisation: Sensitivity analyses

We perform two robustness test to our framework: counting the number of inversions generated by the use of the different templates; and computing the variability of the security index with imperfect knowledge as evidence of internal generalisation.

*Inversion counting.* The Hard/Easy/Default templates show the maximum range of values for our security index per campaign. By looking at Fig. 10, it is natural to ask *how much* these template choices reorder our campaigns w.r.t. said index value. To quantify this, we compute the *inversion count* between any two template-induced campaign rankings, i.e., counting how many pairs of campaigns have their relative order flipped when comparing two rankings. Given the set of $n$ campaigns $\mathcal{C} = \{C_1, \ldots, C_n\}$, we let $\sigma\colon \{1, \ldots, n\} \to \mathcal{C}$, and $\pi\colon \{1, \ldots, n\} \to \mathcal{C}$, be the two permutations encoding the campaign orderings resulting from two different template-induced security indices. Here,

- $\sigma(i) = C_j$ means that campaign $C_j$ occupies position $i$ in the $\sigma$-ranking,
- $\pi(i) = C_k$ means that campaign $C_k$ occupies position $i$ in the $\pi$-ranking,
- $\sigma^{-1}(C_j)$ (resp. $\pi^{-1}(C_j)$) gives the index $i$ such that $\sigma(i) = C_j$ (resp. $\pi(i) = C_j$).

These inverses let us compare the relative positions of any two campaigns across two rankings via the *inversion count*, defined as follows:

$$\mathrm{Inv}(\sigma, \pi) \;=\; \sum_{1 \le i < j \le n} \mathbf{1}\Big[\sigma^{-1}(C_i) < \sigma^{-1}(C_j) \;\wedge\; \pi^{-1}(C_i) > \pi^{-1}(C_j)\Big] \tag{4}$$

i.e., in ranking $\sigma$, campaign $C_i$ comes before $C_j$ and in ranking $\pi$, campaign $C_i$ comes after $C_j$, and where $\mathbf{1}[\cdot]$ is again the indicator function (as in Eq. (1)). This sum enumerates exactly those campaign pairs whose relative order is flipped between the two rankings: e.g., between Hard and Default we count 37 inversions. Table 4 (a) gives the triangular matrix of inversion counts between each pair of the Hard, Easy, and Default template-induced campaign rankings. Table 4 (b) represents the same value as percentages of the max. possible inversions.

Table 4. Pairwise inversion counts between campaign rankings induced by the security indices of Hard, Easy, and Default templates (left); also expressed as percentages of the maximum possible inversions (right). In Table 4b, darker-shaded cells indicate larger inversion percentages.

| | Hard | Easy | Default |
|---|---|---|---|
| Hard | 0 | 50 | 37 |
| Easy | 50 | 0 | 35 |
| Default | 37 | 35 | 0 |

(a) Inversion counts

| | Hard | Easy | Default |
|---|---|---|---|
| Hard | 0.0% | 19.8% | 14.6% |
| Easy | 19.8% | 0.0% | 13.8% |
| Default | 14.6% | 13.8% | 0.0% |

(b) Percent of max. inversions ($\binom{23}{2} = 253$ pairs)

*Imperfect knowledge.* In Fig. 10, the probabilities used to compute the security indices of the different templates where estimated using eq. (1) on the complete dataset of MITRE ATT&CK campaigns. That represents a *perfect knowledge* situation, where the complete history of known attacks is used to estimate the frequency with which a technique is selected to achieve a tactic.

While it is natural to use all available information to build such knowledge base, future attacks are bound to affect these values. It is thus relevant to know how resilient is the security index to such information increase, as the known evidence expands with more campaigns and cybersecurity intelligence gathered by MITRE in time.

To illustrate how this progressive knowledge increase affects our framework, Fig. 11 shows the security index in varying situations of *imperfect knowledge*. There, the index for each campaign is computed from minimal partial knowledge (as soon as the campaign was recorded) to perfect

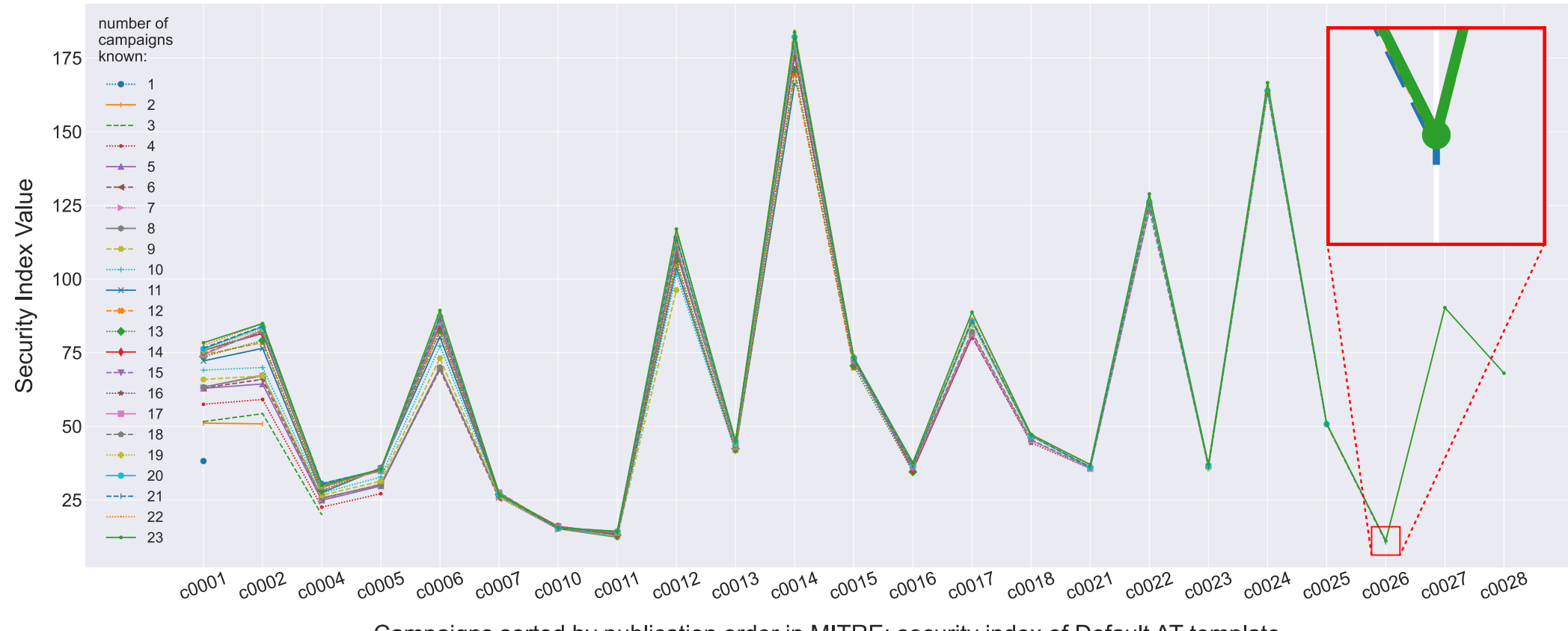

Fig. 11. Variability of the security index with imperfect knowledge: using only the first campaign for techniques frequency computation; only the first two; only the first three; . . .; up to all campaigns (as in Fig. 10)

knowledge (complete dataset). For example, campaign c0026 can be studied in three different cases, as shown in the magnified image that telescopes to the upper-right corner of Fig. 11: *(i)* as soon as it occurred, giving us its collection of techniques to count—dashed blue line, number 21 in the legend to the left of the plot; *(ii)* after campaign c0027 was recorded—dotted orange line, number 22 in the legend; and *(iii)* with present perfect knowledge after campaign c0028—solid green line. Therefore, there are three possible different values for campaign c0026 depicted in Fig. 11: *(i)* 10.81; *(ii)* 11.13; and *(iii)* 11.17. Due to their proximity, these values cannot even be distinguished in the upper-right magnification of Fig. 11: we consider such differences insignificant.

All in all, we observe that the security index is quite robust to computations on partial knowledge scenarios, with variabilities typically below 10% of its absolute value. Some higher variations are observed to the left of the plot, namely for the first two to five campaigns. These jumps in the index correspond to very small datasets, with one (leftmost individual blue dot), two (orange solid line), three (dashed green line), or four campaigns only (dotted red line).

The jumps show how data-based approaches can suffer from non-trivial variability when operating on small datasets such as the aforementioned. Replacing frequentists approaches like eq. (1) for more robust statistical machinery, e.g. Bayesian inference, can help mitigate these variations. That said, we observe that differences are negligible after campaign c0018. Given that the current dataset has eight campaigns on top of that one, we expect that newly discovered campaigns will have only a limited effect on future computations. We note that Fig. 11 is plotted for the Default templates (only); a similar situation can be seen when using the Easy and Hard templates.

## 8 CONCLUSIONS

In this paper we equipped cybersecurity practitioners with tools to perform data-driven comparisons of cybersecurity attack campaigns. This procedure is grounded in MITRE ATT&CK, the *de facto* standard knowledge base for recording cyberattacks happening in the wild. Using our framework, practitioners can extract likelihood data for campaigns and model them in an automatic fashion, via the use of formal AT templates adjustable by the desired interpretation of MITRE data w.r.t. the expected difficulty of an attack.

Thus, this framework is able to compare MITRE ATT&CK campaigns quantitatively and automatically, by introducing:

(1) a templetised method to construct attack tree models out of Mitre data, that avoids time-consuming and bias-prone manual modelling;
(2) a quantification heuristic—also based on Mitre data—that is compounded via a formal logic into a security index, indicative of the likelihood of observing a campaign succeed according to past evidence;
(3) a comparison of the security index of any two Mitre ATT&CK campaigns, as well as validation of manually-constructed AT models for them, including an expected *security range* in which a model representing Mitre data should lie.

*Reflection, including threats to validity.* In light of our analysis, we briefly reflect on validity and validation of the proposed methodology. Firstly, Mitre campaign records constitute the link of our methodology to reality. This is clearly reflected e.g., on security indices in Table 3, since these quantities are computed via a process that is directly based on data retrieved from Mitre ATT&CK. Theoretically, the feasible range of this security index is any positive real number. However, in practice we observe that Wocao is the hardest campaign, whose hard template yields a security index around 300. This suggests that the practical range of (necessarily positive) security indices, when considering realistic real-world cyberattack campaigns, should typically not surpass the 1k mark. However, we highlight once again that these absolute numbers are of little use in and on themselves, but instead useful to compare and prioritise which campaigns/tactics/techniques to mitigate first, namely those that correspond to situations of lower security.

We give a concrete security scenario example by considering the indices for Wocao and Dream Job. Results in Table 3 suggest that, when operating under a limited budget, one should prioritize the mitigation of tactics and techniques included in the Dream Job cyberattack. The advice is based on the security index of Dream Job being lower than Wocao: while the difference favours the former for the Easy template—when security considerations are most lax—such difference is only 1.97 points. In contrast, when the security setting is default or hardened, the difference is resp. 55.06 or 125.35 points against Dream Job. This suggests that attacks using the approach of Dream Job, if applicable to the system deployment under consideration, are considerable more likely to succeed than attacks that follow the killchain of Wocao.

Further abstractions and assumptions operated in modelling and computing quantities from Mitre ATT&CK data are transparently highlighted in this paper: this transparency is intended to help practitioners interpreting and validating proposed results obtained via our methodology. Moreover, if one needs to modify these assumptions, our white-box method allows practitioners to tailor this framework to their specific needs. Ultimately, we envision our methodology and framework to be used in live scenarios, to help developing and validating the modelling of a campaign at the same time that its technical report is prepared for compatibility with Mitre ATT&CK.

A corollary of the above and a threat to validity is that, if field experts–recording a campaign into Mitre–model a cyberattack in Mitre ATT&CK such that it does not reflect what the technical report reads, then our method would produce an AT that does not resemble the truth, but rather the data interpreted from Mitre incorrectly. This is a direct consequence of adhering to data proposed by the Mitre knowledge base, whose correct encoding (e.g., in the Mitre ATT&CK Navigator) falls on practitioners' recording procedures.

Another limitation of this work comes from the expert validation, which is relatively small-scale as the expert-constructed trees have been provided for two campaigns only. For these campaigns, we provide the detailed methodology followed to reach the custom ATs in Appendix A and Appendix B. However, it remains challenging to assess how robust the expert agreement is really and whether the results generalize beyond the two selected examples. To guarantee broader applicability, further validation with cybersecurity experts is needed.

Finally, a further limitation is the need to operate with imperfect knowledge regarding the attack campaigns, which impacts the representativeness of the likelihood values carved out of MITRE data. There is, however, no better alternative: as discussed in Sec. 2.2, the data gathered in MITRE ATT&CK—the authoritative source for cybersecurity intelligence—is the closest approximation to the knowledge available to and harvested by field experts. Furthermore, the need to operate with imperfect information is a common hurdle in cybersecurity practice.

*Future work.* Our work opens several interesting directions for future research. First, one might extend this framework to reason about other quantitative security metrics. E.g. by introducing severity of attacks and adopting a similar data-driven approach grounded in other knowledge bases, such as the Common Vulnerability Scoring System (CVSS) by NIST, or the Exploit Prediction Scoring System (EPSS). For this, a first resource to consider are cyberintelligence statistial reports, to distinguish between potential and effectively-used attacks paths [7]. Such reports could also serve for external validation, as opposed to the internal-generalisation studies of Sec. 7.3, by providing out-of-distribution points to compare the computed probabilities against.

Secondly, one could extend the proposed framework and introduce defence mechanisms, exploiting richer modelling languages – such as attack–defence trees [20] and attack graphs [17]. Lastly, support for Pareto optimal solutions [3] w.r.t. different metrics could further propel quantitative decision making, and help in considering trade-offs, e.g. between cost of defences and probability of successful attacks [26]. This is not yet present in ATM [34], which would need to be extended as well. In general, it would be interesting to explore what other functionality of ATM would be useful in assessing MITRE campaigns. In this regard, cATM provides one approach, but a more thorough comparison between ATM's functionality and cybersecurity practicioners' could yield more powerful tools.

## ACKNOWLEDGMENTS

This work was partially supported by the European Union under NextGenerationEU projects D53D23008400006 *Smartitude* funded by MUR under the PRIN 2022 program, and PE00000014 CUP E63C24000590001 *SERICS (COVERT)* funded by MUR under the National Recovery and Resilience Plan; MSCA grants 101008233 *MISSION* and 101067199 *ProSVED*, the ERC Consolidator Grant 864075 *CAESAR*, the ERC Proof of Concept grant 101187945 *RUBICON*, and the NWO grant NWA.1160.18.238 *PrimaVera* and grant KIC1.VE01.20.004 *HEWSTI*. Views and opinions expressed are however those of the author(s) only and do not necessarily reflect those of the European Union, or The European Research Executive Agency, or NWO. Neither the European Union nor the granting authorities can be held responsible for them.

*Data and Artifact Availability.* We provide public access to all data and programs created and used in this work at DOI `10.5281/zenodo.14193936`.

## A EXPERT-DRIVEN ATTACK TREE FOR THE WOCAO CAMPAIGN

The *ad hoc* model of the Wocao campaign introduced in Sec. 3.2 is based on the publicly available report at [46]. In interpreting the technical information of the report, the following thirteen expert-driven decisions were made:

(1) We further refined MITRE technique *Exploit Public-Facing Application* (`T1190`) by searching vulnerability databases and matching two possible CVEs that attackers could have exploited to leverage vulnerabilities in JBoss webservers. These are CVE-2017-12149 and CVE-2017-7504[¶]: probability values are then attributed via the Exploit Prediction Scoring System (EPSS, https://www.first.org/epss/).

[¶]See respectively https://nvd.nist.gov/vuln/detail/CVE-2017-12149 and https://nvd.nist.gov/vuln/detail/CVE-2017-7504.

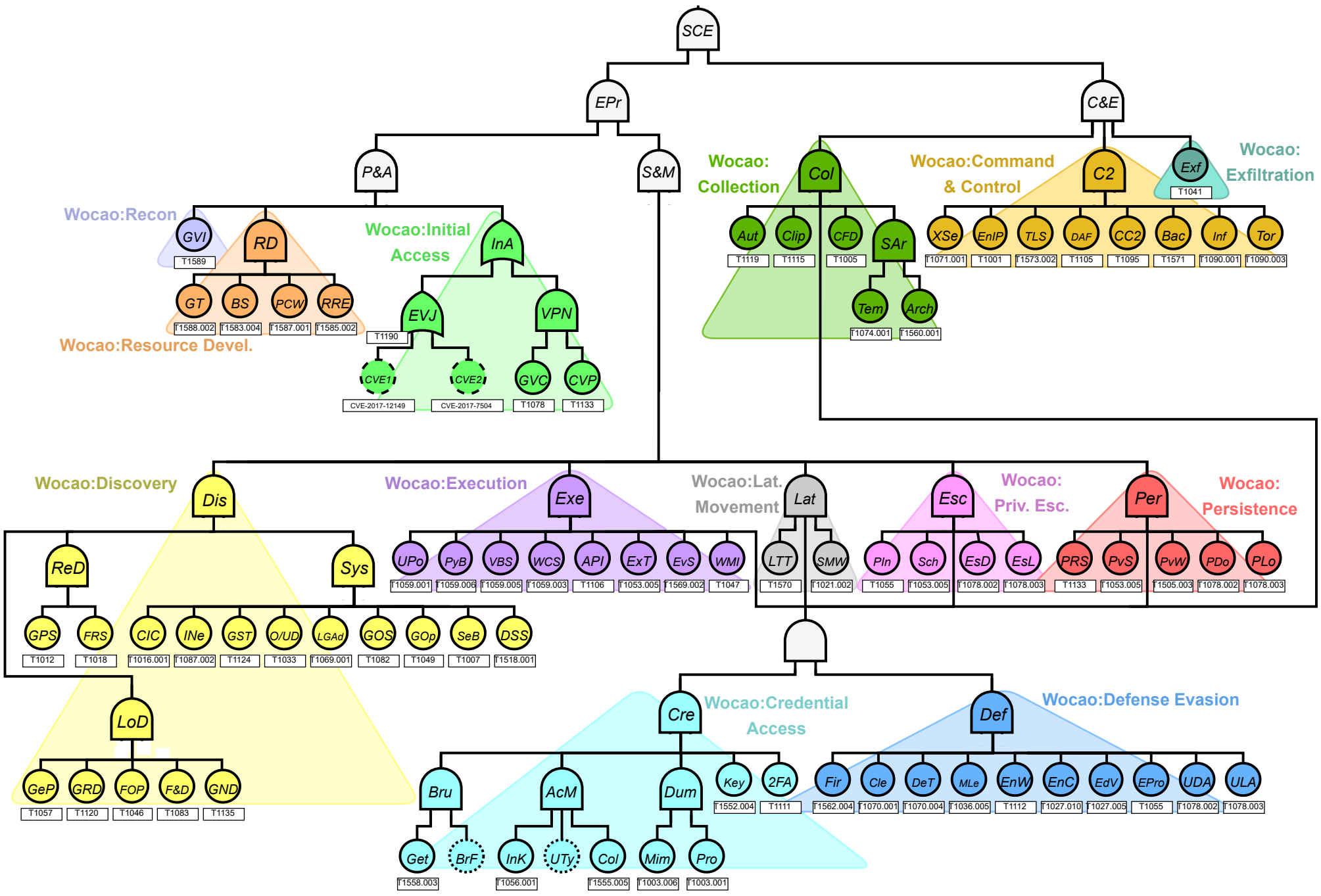

Fig. 12. AT custom model for the Wocao campaign from MITRE [https://attack.mitre.org/versions/v14/campaigns/C0014/]. This is the same AT model shown in Fig. 5 on page 11. The meaning of the abbreviations in the figure can be found in Table 1 on page 12.

(2) The *Execution* tactic includes the *Visual Basic* sub-technique (`T1059.005`), with a comment that reads "Threat actors used VBScript to conduct reconnaissance on targeted systems"; However, `T1059.005` does not appear to be listed under the *Reconnaissance* tactic:
   - `T1059.005`, which involves the execution of a script, was modelled as a BAS under the *Execution* tactic and not under *Reconnaissance*, thus reflecting MITRE data.

(3) Tactic *Initial Access* includes the technique *Valid Accounts* (`T1078`), as well as its sub-techniques *Domain Accounts* (`T1078.002`) and *Local Accounts* (`T1078.003`); However, a technical comment states that these credentials "found during intrusion" were used for "lateral movement and privilege escalation", and the MITRE tactics *Lateral Movement* and *Privilege Escalation* do list the sub-techniques as part of their steps:
   - For tactic *Initial Access* only the parent technique `T1078` was kept, because MITRE does not mention which specific sub-technique is needed for initial access, but obtaining (any) valid VPN credential is a known requirement [46].

(4) An analogous rationale was applied to tactics *Privilege Escalation*, *Defense Evasion*, and *Persistence*, for which MITRE lists *Domain Accounts* (`T1078.002`) and *Local Accounts* (`T1078.003`) as sub-techniques, following the comment mentioned in the previous item:
   - Sub-techniques `T1078.002` and `T1078.003` were modelled as (individual, see next item) BASes under the sub-trees corresponding to these tactics.

(5) In MITRE some techniques are shared among tactics, e.g. *Scheduled Tasks* (`T1053.005`) is listed under the *Execution*, *Persistence*, and *Privilege Escalation* tactics; However, while sharing the technique ID is logical from an archiving perspective, its empirical application is typically different for each tactic—see e.g. the case of `T1078.002` and `T1078.003` above, or compare scheduling a task to execute a recurring command vs. using a task scheduler to gain privileges, or to persist in the system by launching a web shell.

- These techniques are thus modelled as independent in the AT, e.g. there are three `T1053.005` BASes, each under a different tactic and with a unique name.

(6) The *Keylogging* technique (`T1056.001`) was used for tactics *Collection* and *Credential Access*; However, as per the technical report, Wocao was a cyber-espionage operation where undetected data collection—including credentials—is the overall attack objective:

- Instead of a BAS with `T1056.001`, the sub-trees of tactics *Defense Evasion* and *Credential Access*—which include `T1056.001`—were modelled as children of the *Collection* tactic, to reflect in greater detail the multiple types of data collected.

(7) The same rationale was applied to the tactics *Execute*, *Lateral Movement*, *Privilege Escalation*, and *Persistence*—because, in every one of those tactics, the steps to evade defenses and collect credentials are an integral part of the attacker's maneuvers:

- The sub-trees of tactics *Credential Access* and *Defense Evasion* were modelled as children of all the aforementioned tactics.

(8) Tactics in MITRE are listed in a loose order, e.g. *Discovery*—that includes gathering data on running processes (`T1057`)—is the ninth tactic, even though it may be needed to discover running processes first, to later perform the required *Execution* (fourth tactic):

- In terms of the top-level gate, the sub-tree corresponding to the *Discovery* tactic was modelled between tactics *Initial Access* and *Execution*, because in the Wocao campaign it is necessary to perform technique `T1057`, from the *Discovery* tactic, before being able to perform technique `T1055` (*Process injection* from the *Execution* tactic).

(9) For this campaign, there are no techniques pertaining to the *Impact* tactic listed in MITRE:

- The *Impact* sub-tree was not modelled, which in any case is not essential to assess the probability of success of this attack.

(10) Furthermore, we introduce two custom leaves attributed with probabilities from domain experts in the sub-tree for tactic *Credential Access* (dotted border). These represent respectively the probability of successfully bruteforcing passwords – see leaf with the *BrF* acronym – and the probability that a user types his master password – *UTy*. We opted to add these leaves as they model steps that were explicitly referenced in the technical report [46] and showcase an exemplar situation of AT nodes that might require expert-guided likelihood attribution depending on the specific setting at hand.

Finally, considering data granularity (see e.g. Sec. 2.2.3), we adopted the following strategies when faced with mixed data:

(11) In the *Discovery* tactic we eliminate technique `T1016` *System Network Configuration* in favour of the more fine-grained information provided by the sub-technique `T1016.001` *Internet Connection Discovery*. Furthermore, we eliminate technique `T1518` *Software Discovery* in favour of the more fine-grained information provided by the sub-technique `T1518.001` *Security Software Discovery*.

(12) In the *Command and Control* tactic we eliminate technique `T1090` *Proxy* in favour of the more fine-grained information provided by the two sub-techniques `T1090.001` *Internal Proxy* and `T1090.003` *Multi-hop Proxy*.

(13) Further incoherence given by the simultaneous presence of `T1078` *Valid Accounts* and its sub-techniques was already ruled out by choices presented previously.

## B EXPERT-DRIVEN ATTACK TREE FOR THE DREAM JOB CAMPAIGN

The *ad hoc* model of the Dream Job campaign introduced in Sec. 3.3 represents the espionage operation described in the technical report [8]. Here we present the complete AT model.

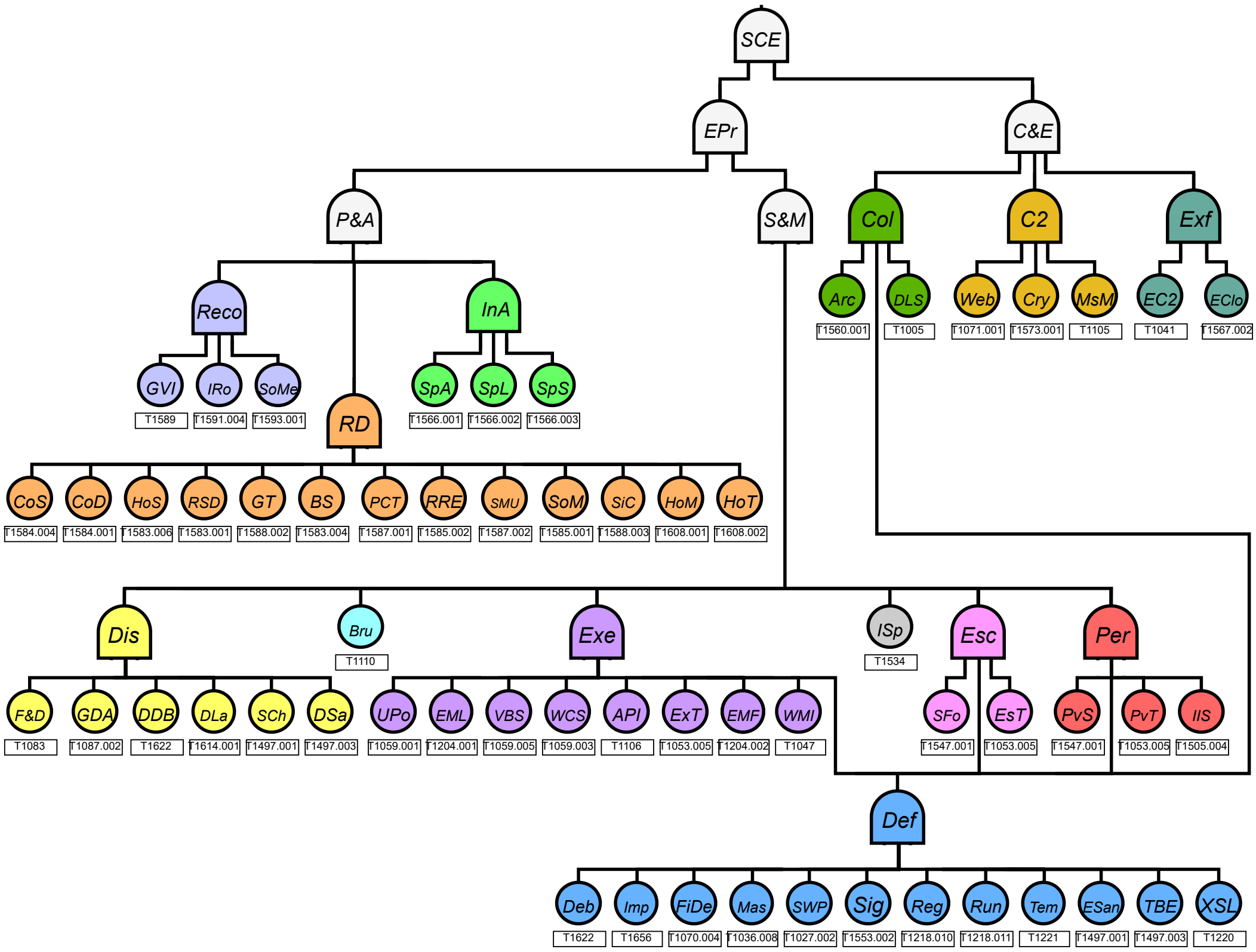

Fig. 13. AT custom model for the Dream Job MITRE campaign [https://attack.mitre.org/versions/v14/campaigns/C0022/]. Acronyms for gates in the AT can be found on page 35.

Table 5. Abbreviations for the AT in Fig. 13.

| **Operation Dream Job TLE:** | | **Discovery:** | | **Defense Evasion:** | |
|---|---|---|---|---|---|
| Success in Cyber Espionage | SCE | Discovery | Dis | Defense Evasion | Def |
| Establish Presence | EPr | Files & Dirs Scan | F&D | Debugger Evasion | Deb |
| Collect & Exfiltrate | C&E | Get Domain Accounts | GDA | Impersonation | Imp |
| Prep. & Access | P&A | Detect Debuggers | DDB | File Deletion | FiDe |
| Spread & Maintain | S&M | Detect System Language | DLa | File Type Masquerade | Mas |
| **Reconnaissance:** | | System Checks | SCh | Software Packing | SWP |
| Reconnaissance | Reco | Detect Sandboxes | DSa | Code Signing | Sig |
| Gather Victim Info | GVI | **Execution:** | | Use Regsvr32 | Reg |
| Identify Roles | IRo | Execution | Exe | XSL Script Processing | XSL |
| Social Media | SoMe | Windows Command Shell | WCS | Use Rundll32 | Run |
| **Resource Devel.:** | | Execute via WMI | WMI | Template Injection | Tem |
| Resource Devel. | RD | VB Malicious Macro | VBS | Evade Sandboxes | ESan |
| Get Tools | GT | Use Powershell | UPo | Time-Based Evasion | TBE |
| Buy Server | BS | Obtain User-Agent via API | API | **Collection:** | |
| Prepare Custom Tools | PCT | Exec with Scheduled Tasks | ExT | Collection | Col |
| Register Rogue Email Addr. | RRE | Exec. via Malicious File | EMF | Archive via Utility | Arc |
| Register Same Domain | RSD | Exec. via Malicious Link | EML | Data from Local Sys. | DLS |
| Hosting Services | HoS | **Credential Access:** | | **Command & Control:** | |
| Compromise Domains | CoD | Brute Force Attacks | Bru | Command & Control | C2 |
| Compromise Servers | CoS | **Lateral Movement:** | | Web Protocols | Web |
| Sign Malware & Utils | SMU | Internal Spearphish. | ISp | Symmetric Cryptography | Cry |
| Social Media | SoM | **Privilege Escalation:** | | Multistage Malware Ingress | MsM |
| Signing Certificates | SiC | Privilege Escalation | Esc | **Exfiltration:** | |
| Host Malware | HoM | Escalate via Startup Folder | SFo | Exfiltration | Exf |
| Host Tools | HoT | Escalate via Task Scheduler | EsT | Exfil over C2 Channel | EC2 |
| **Initial Access:** | | **Persistence:** | | Exfil over Cloud | EClo |
| Initial Access | InA | Persistence | Per | | |
| Spearphish. (Attachment) | SpA | Persist via Startup | PvS | | |
| Spearphish. (Link) | SpL | Persist via Task Scheduler | PvT | | |
| Spearphish. (Service) | SpS | IIS Components | IIS | | |

## C SEMANTICS OF CATM

Layer 1 formulae are evaluated on a pair $(T, A)$ of an AT $T$ and an attack $A$ on $T$:

$$\begin{aligned} (A,T) &\models e && \text{iff } f_T(e,A) = 1, \\ (A,T) &\models \neg\phi && \text{iff } A, T \not\models \phi, \\ (A,T) &\models \phi \wedge \phi' && \text{iff } A, T \models \phi \text{ and } A, T \models \phi'. \end{aligned}$$

For layer 2 formulae we need a bit more notation. Given an attack tree $T$, for a layer 1 formulae $\phi$ and an attack $A$, we define the set of *successful attack* on $\phi$ to be

$$\mathrm{Succ}(\phi) = \{A \in \mathscr{A} \mid (A,T) \models \phi\}.$$

We then define the *minimal attacks* on $\phi$ to be

$$[\![\phi]\!] = \{A \in \mathrm{Succ}(\phi) \mid \forall A' \in \mathrm{Succ}(\phi) : A' \not\subset A\}\,.$$

Now let an interval-attributed tree $\mathsf{T} = (T, \mathscr{L}, \boldsymbol{\iota})$, with $\mathscr{L} = \{L_1, \ldots, L_l\}$; so $\boldsymbol{\iota} = (\iota_1, \ldots, \iota_l)$. For $i \leq l$, the *metric value of* $\phi$ is defined as

$$\breve{\iota}_i(\phi) = \left[ \bigtriangledown_{A \in [\![\phi]\!]} \bigtriangleup_{a \in A} \iota_i^-(a),\ \bigtriangledown_{A \in [\![T]\!]} \bigtriangleup_{a \in A} \iota_i^+(a) \right];$$

this is an interval in the ordered semiring $L_i$. For an interval-attributed tree $\mathsf{T}$, we now define the semantics for layer 2 formulæ as follows:

$$\mathsf{Val}_{\mathsf{T}}(\mathbb{V}_k(\phi)) = \check{\iota}_k(\phi)$$
$$\mathsf{Val}_{\mathsf{T}}(\psi[a \xmapsto{k} [L_a, U_a]]) = \mathsf{Val}_{\mathsf{T}(\mathfrak{i}[\iota_k(a) \xmapsto{k} [L_a, U_a]])}(\psi),$$

where $\mathfrak{i}[\iota_k(a) \xmapsto{k} [L_a, U_a]]$ stands for the interval attribution obtained by replacing $[\iota_k^-(a), \iota_k^+(a)]$ in $\mathfrak{i}$ by $[L_a, U_a]$.